\documentclass[11pt]{article}
\usepackage[usenames,dvipsnames,svgnames,table]{xcolor} 
\usepackage[obeyspaces,hyphens,spaces]{url}
\usepackage{jcapmod}
\usepackage{geometry}
\usepackage{changepage}

\usepackage{tabu}
\usepackage{booktabs}
\usepackage[english]{babel}
\usepackage{amsmath, amssymb, amsbsy, amstext, amsthm, simplewick}
\usepackage{hyperref}
\usepackage{graphicx}
\usepackage{amsfonts}
\usepackage{enumitem}
\usepackage{upgreek}
\usepackage{framed}
\usepackage{tensor}
\usepackage{pifont}
\usepackage{latexsym, mathrsfs}
\usepackage{array}
\usepackage{hyperref}
\usepackage{xspace}
\usepackage{longtable}
\usepackage{multirow}
\usepackage{cases}
\usepackage{empheq}
\usepackage{bm}
\usepackage{bbm}
\usepackage[table]{xcolor}
\usepackage[dvipsnames]{xcolor}
\usepackage{colortbl}
\usepackage{tablefootnote}
\usepackage{braket}
\usepackage[utf8]{inputenc}

\usepackage[load-configurations=astronomy, range-units=brackets, range-phrase=-, per-mode=reciprocal, mode=math]{siunitx}
\usepackage{threeparttable}
\usepackage{subfig}

\usepackage{ragged2e}

\usepackage[normalem]{ulem}

\usepackage{hhline}
\usepackage{array}
\newcolumntype{P}[1]{>{\centering\arraybackslash}p{#1}}
\usepackage{booktabs}
\usepackage{tikz}
\usetikzlibrary{calc,shadings,patterns,tikzmark,fadings}

\definecolor{Blue}{rgb}{0.25, 0.41, 0.88}
\definecolor{Red}{rgb}{0.92,0.,0.}
\definecolor{darkorange}{rgb}{1.0,0.549,0.}
\definecolor{cobalt}{RGB}{44, 98, 120}
\definecolor{Mathematica1}{rgb}{0.368417, 0.506779, 0.709798}
\definecolor{Mathematica2}{rgb}{0.880722, 0.611041, 0.142051}
\definecolor{Mathematica3}{rgb}{0.560181, 0.691569, 0.194885}
\definecolor{Mathematica4}{rgb}{0.922526, 0.385626, 0.209179}
\definecolor{Mathematica5}{rgb}{0.528488, 0.470624, 0.701351}
\definecolor{Mathematica6}{rgb}{0.772079, 0.431554, 0.102387}
\definecolor{Mathematica7}{rgb}{0.363898, 0.618501, 0.782349}
\definecolor{Mathematica8}{rgb}{1, 0.75, 0}
\definecolor{Mathematica9}{rgb}{0.647624, 0.37816, 0.614037}
\definecolor{plotBlue}{RGB}{94, 130, 181}
\definecolor{plotRed}{RGB}{233, 85, 54}
\definecolor{plotGreen}{RGB}{142, 176, 50}
\definecolor{plotPurple}{RGB}{135, 120, 178}

\renewcommand\_{\textunderscore\allowbreak}

\definecolor{cornellRed}{HTML}{B31B1B}
\definecolor{cornellBlue}{HTML}{0068AC}
\definecolor{cornellGreen}{HTML}{6EB43F}

\newcolumntype{C}[1]{>{\centering\let\newline\\\arraybackslash\hspace{0pt}}m{#1}}

\def\d{{\rm d}}

\newcommand{\affsize}{10}
\newcommand{\affvskip}{\vskip 1pt}

\makeatletter
\newlength{\apb@width}
\newcommand{\autoparbox}[2][c]{\settowidth{\apb@width}{#2}\parbox[#1]{\apb@width}{#2}}

\makeatother

\numberwithin{equation}{section}

\def\beq{\begin{equation}}
\def\eeq{\end{equation}}

\def\bea{\begin{eqnarray}}
\def\eea{\end{eqnarray}}

\def\d{{\rm d}}

\def\beq{\begin{equation}}
\def\eeq{\end{equation}}
\def\bea{\begin{eqnarray}}
\def\eea{\end{eqnarray}}

\def\d{{\rm d}}

\def\d{{\rm d}}

\DeclareRobustCommand{\SkipTocEntry}[4]{}

\usepackage{colortbl}
\definecolor{blue2}{cmyk}{1, 0.1, 0.1, 0}

\definecolor{pyBlue}{RGB}{31, 119, 180}
\definecolor{pyRed}{RGB}{214, 39, 40}
\definecolor{pyGreen}{RGB}{44, 160, 44}
\definecolor{pyBlue2}{RGB}{0, 111, 237}
\definecolor{pyRed2}{RGB}{224, 52, 36}

\makeatletter
\def\Ddots{\mathinner{\mkern1mu\raise\p@
\vbox{\kern7\p@\hbox{.}}\mkern2mu
\raise4\p@\hbox{.}\mkern2mu\raise7\p@\hbox{.}\mkern1mu}}
\makeatother

\begin{document}

\pagenumbering{roman}

\begin{titlepage}

\newgeometry{vmargin={15mm}, hmargin={16mm,16mm}}

\baselineskip=15.5pt \thispagestyle{empty}
\begin{flushright}

\end{flushright}
\vspace{0cm}

\begin{center}
\begin{minipage}{1.0\textwidth}
\centering
{\fontsize{20}{0}\selectfont \bfseries  In Pursuit of Love:} \\[12pt] 

{\fontsize{15.5}{0}\selectfont \bfseries {First Templated Search for Compact Objects with \\[6pt] Large Tidal  Deformabilities in the LIGO-Virgo Data}}
 
\end{minipage}
\end{center}


\vspace{0.001cm}

\begin{center}
\begin{minipage}{1\textwidth}
\centering
{\fontsize{13}{0} \selectfont Horng Sheng Chia$^{1}$, Thomas D. P. Edwards$^{2}$, Digvijay Wadekar$^{1}$, Aaron Zimmerman$^{3}$,} \\[4pt] {\fontsize{13}{0} \selectfont Seth Olsen$^{4}$, Javier Roulet$^{5}$, Tejaswi Venumadhav$^{6,7}$, Barak Zackay$^{8}$, and Matias Zaldarriaga$^{1}$} 
\end{minipage}
\end{center}


\begin{center}
\begin{minipage}[c]{1.0\textwidth}
\centering

\textsl{\fontsize{\affsize}{0}\selectfont $^1$ School of Natural Sciences, Institute for Advanced Study, Princeton, NJ 08540, USA}

\affvskip

\textsl{\fontsize{\affsize}{0}\selectfont $^2$ Department of Physics and Astronomy, Johns Hopkins University, Baltimore, Maryland 21218, USA}

\affvskip

\textsl{\fontsize{\affsize}{0}\selectfont $^3$ Center for Gravitational Physics, University of Texas at Austin, Austin, Texas 78712, USA}

\affvskip

\textsl{\fontsize{\affsize}{0}\selectfont $^4$ Department of Physics, Princeton University, Princeton, NJ 08540, USA}

\affvskip

\textsl{\fontsize{\affsize}{0}\selectfont $^5$ TAPIR, Walter Burke Institute for Theoretical Physics, California Institute of Technology, Pasadena, CA 91125, USA}

\affvskip

\textsl{\fontsize{\affsize}{0}\selectfont $^6$ Department of Physics, University of California at Santa Barbara, Santa Barbara, California 93106, USA}

\affvskip

\textsl{\fontsize{\affsize}{0}\selectfont $^7$ International Centre for Theoretical Sciences, Tata Institute of Fundamental Research, Bangalore 560089, India}

\affvskip

\textsl{\fontsize{\affsize}{0}\selectfont$^8$ Department of Particle Physics \& Astrophysics, Weizmann Institute of Science, Rehovot 76100, Israel}

\end{minipage}
\end{center}

\vspace{0.3cm}

\begin{center}
\begin{minipage}{0.88\textwidth}
\hrule \vspace{10pt}
\noindent {\bf Abstract}\\[0.1cm]
We report results on the first matched-filtering search for binaries with compact objects having large tidal deformabilities in the LIGO-Virgo gravitational wave (GW) data. The tidal deformability of a body is quantified by the ``Love number" $\Lambda \propto \hskip 1pt (r/m)^5$, where $r/m$ is the body's (inverse) compactness. Due to its strong dependence on compactness, the $\Lambda$ of larger-sized compact objects can easily be many orders of magnitude greater than those of black holes and neutron stars, leaving phase shifts which are sufficiently large for these binaries to be missed by binary black hole (BBH) templated searches. 
In this paper, we conduct a search using inspiral-only waveforms with zero spins but finite tides, with the search space covering chirp masses $3 M_\odot < \mathcal{M} < 15 M_\odot$ and effective tidal deformabilities $10^2 \lesssim \tilde{\Lambda} \lesssim 10^6$. We find no statistically significant GW candidates. This null detection implies an upper limit on the merger rate of such binaries in the range $[1-300] \hskip 2pt \text{Gpc}^{-3} \text{year}^{-1}$, depending on $\mathcal{M}$ and $\tilde{\Lambda}$. While our constraints are model agnostic, we discuss the implications on beyond the Standard Model scenarios that give rise to boson stars and superradiant clouds. Using inspiral-only waveforms we recover many of the BBH signals which were previously identified with full inspiral-merger-ringdown templates.
We also constrain the Love number of black holes to $\Lambda \lesssim 10^3$ at the 90\% credible interval. Our work is the first-ever dedicated template-based search for compact objects that are not only black holes and neutron stars. Additionally, our work demonstrates a novel way of finding new physics in GW data, widening the scope of potential discovery to previously unexplored parameter space.

\vskip15pt
\hrule
\vskip10pt
\end{minipage}
\end{center}

\restoregeometry

\end{titlepage}

\thispagestyle{empty}
\setcounter{page}{2}
\tableofcontents

\newpage
\pagenumbering{arabic}
\setcounter{page}{1}

\clearpage
\section{Introduction}
 \label{sec:introduction}

All gravitational wave (GW) matched-filtering searches to date have been performed using template banks constructed with aligned-spin binary black hole (BBH) waveforms; see e.g. Refs.~\cite{LIGOScientific:2018mvr, LIGOScientific:2020ibl, LIGOScientific:2021usb, LIGOScientific:2021djp, Venumadhav:2019tad, Venumadhav:2019lyq, Olsen:2022pin, Ajit-O3b, Nitz:2018imz, Nitz:2019hdf, Nitz:2021uxj, Nitz:2021zwj}. Although matched-filtering is the optimal linear filter in stationary Gaussian noise~\cite{matched_filter_original, Maggiore:2007ulw}, it relies on precise phase coherence between the template and the signal~\cite{Sathyaprakash:1991mt, Dhurandhar:1992mw}. This sensitivity to phase coherence significantly diminishes our ability to detect putative new signals that differ from those contained within the BBH template banks~\cite{Chia:2020psj}, thereby limiting the scope of potential new discoveries. In this paper, we expand the space of templates and search for a wider class of compact objects in the current public LIGO-Virgo data. 

\vskip 4pt

The phase of the GWs emitted by a binary system is extremely sensitive to the physics of the individual compact objects~\cite{Cutler:1992tc}. 
This encoded physics is especially interpretable in the inspiral regime of the binary coalescence, where the binary components are well separated and perturbative corrections to the orbital dynamics in the form of the post-Newtonian (PN) expansion can be derived analytically~\cite{Blanchet:2013haa, Porto:2016pyg, Levi:2018nxp, Bern:2021dqo}. At leading-order, the orbiting bodies can be modeled as point particles which are uniquely characterized by their masses and spins. However, as the binary approaches merger, various effects associated with the \textit{finite size} of the bodies can provide important contributions to the phase evolution of the waveform. For example, for a spinning body the leading-order finite-size effect is the spin-induced quadrupole, which first appears at 2PN order~\cite{Thorne:1980ru, Poisson:1997ha}. Though formally the leading finite-size contribution, the effect of this term on the GW signal is suppressed if the object's dimensionless spin is small~\cite{Chia:2020psj, Chia:2022rwc}. This is typically the case for objects with large sizes, since their spins are bounded by the mass-shedding upper limit~\cite{Paschalidis:2016vmz}.

\vskip 4pt

For non-spinning or slowly-spinning compact objects, the tidal deformability of each body provides the dominant finite-size effect on the inspiral~\cite{love, nrgr, poisson_will_2014}. This is a conservative tidal effect whereby the gravitational perturbation sourced by the binary companion changes the multipolar structure of the body. The tidal deformability of a body is quantified by a set of ``Love numbers"~\cite{love}, with the leading quadrupolar Love number first appearing in the waveform at 5PN order~\cite{nrgr, Flanagan:2007ix}. Depending on the nature of the body, the quadrupolar Love number could be large enough to introduce significant changes  to the waveform~\cite{Flanagan:2007ix, Vines:2011ud}. For black holes, the Love number is zero\footnote{
    In fact, the Love numbers of black holes not only vanish for the leading quadrupolar perturbation but also to all orders in the multipole expansion of the external tidal field, for all values of black hole spin, and for both the electric- and magnetic-type perturbations in General Relativity (GR)~\cite{Chia:2020yla, Charalambous:2021mea, Binnington:2009bb, Damour:2009vw, Kol:2011vg}. \label{footnote:bh_zerolove}
}~\cite{Chia:2020yla, Charalambous:2021mea, Kol:2011vg, Binnington:2009bb, Damour:2009vw} and is therefore ignored when building template banks for BBH searches. For solar-mass neutron stars, the Love number is sufficiently close to the black hole value~\cite{TheLIGOScientific:2017qsa, Abbott:2018wiz} that BBH templates are effective at detecting binary neutron star (BNS) systems. 
However, for subsolar-mass neutron stars (with masses $\lesssim 1 M_\odot$)~\cite{Yagi:2013awa, Silva:2016myw} and many beyond the Standard Model (BSM) objects~\cite{Baumann:2018vus, DeLuca:2021ite, Mendes:2016vdr, Sennett:2017etc, Cardoso:2017cfl, Ryan:2022hku, Diedrichs:2023trk, Brito:2015pxa, Jain:2021pnk, Gorghetto:2022sue, Amin:2022pzv} proposed in the literature, the Love numbers can easily be orders of magnitude larger. In these cases, the Love numbers can be large enough for the binary system to be missed by BBH searches even if the signals have detectable signal-to-noise (SNR) ratio in the LIGO-Virgo data; see \S\ref{sec:why_new_search} and Fig.~\ref{fig:effectualness} below.

\subsection{Why is a New Search Necessary?} \label{sec:why_new_search}

In order to understand why astrophysical bodies (other than black holes) naturally have large Love numbers, it is instructive to consider the leading tidal effect, which is the induced quadrupolar deformation~\cite{poisson_will_2014}
\begin{equation}
\delta Q_{ij} = - \Lambda \hskip 1pt m^5 \hskip 1pt  \mathcal{E}_{ij} \, , \qquad \Lambda = \frac{2}{3} \hskip 1pt k \left( \frac{r}{m} \right)^5 \, . \label{eqn:tides}
\end{equation}
Here $\delta Q_{ij}$ is the induced mass quadrupole, $\mathcal{E}_{ij}$ is the external tidal field, $\Lambda$ is the dimensionless quadrupolar Love number, $m$ is the object mass, $r$ is the stellar radius, and $k$ is a dimensionless constant (often called the second Love number) whose value depends on the internal structure of the body. For black holes, $k = 0$~\cite{Chia:2020yla, Charalambous:2021mea, Binnington:2009bb, Damour:2009vw, Kol:2011vg}; see Footnote~\ref{footnote:bh_zerolove}. For neutron stars, depending on the nuclear equation of state, $k$ ranges between $0.1$ and $1$~\cite{Hinderer:2007mb, Binnington:2009bb, Damour:2009vw, Yagi:2013awa}. For compact objects that exist in many BSM scenarios, such as superradiant clouds formed around rotating black holes and boson stars, $k$ can be as large as $\sim 10^3$~\cite{Baumann:2018vus, DeLuca:2021ite, Mendes:2016vdr, Sennett:2017etc, Cardoso:2017cfl, Ryan:2022hku, Diedrichs:2023trk}.\footnote{
    The coefficients $k$ for many exotic compact objects with smoothly-varying density distributions, such as self-gravitating scalar field configurations, are generally not well determined because their ``radii" $r$ are not unambiguously defined. In these cases, it is more appropriate to consider $\Lambda$ as the overall measure of their tidal deformabilities. \label{ref:k2_illdefined} 
} In addition to the dependence on $k$, $\Lambda$ depends sensitively on the body's stellar radius and scales as $\Lambda \propto (r/m)^5$, which can be deduced straightforwardly through dimensional analysis. 
The ratio $r/m$ is often called the (inverse) compactness of the body and can span multiple orders of magnitude depending on the system under consideration. For example, $r/m = 2$ for Schwarzschild black holes; $r/m$ is approximately $2.25 - 10$ for solar-mass neutron stars and $\sim 10-10^2$ for subsolar-mass neutron stars~\cite{Hinderer:2007mb, Binnington:2009bb, Damour:2009vw, Yagi:2013awa}; no such compactness bound exists for BSM compact objects since there are often free parameters associated to the new physics which can accommodate objects with large radii. Due to the sensitive $\Lambda \propto (r/m)^5 $ scaling, $\Lambda$ can be easily enhanced by several orders of magnitude, counterbalancing the suppressive effect of the $v^{10}$ factor in this 5PN phase term~\cite{nrgr, Flanagan:2007ix}. 
Indeed, for sufficiently large values of $\Lambda$, the phase evolution of such binary systems would be significantly different from those of BBH waveforms, potentially resulting in them being missed by BBH searches.

\vskip 4pt

\begin{figure}[t!]
    \centering
    \includegraphics[width=0.8\textwidth]{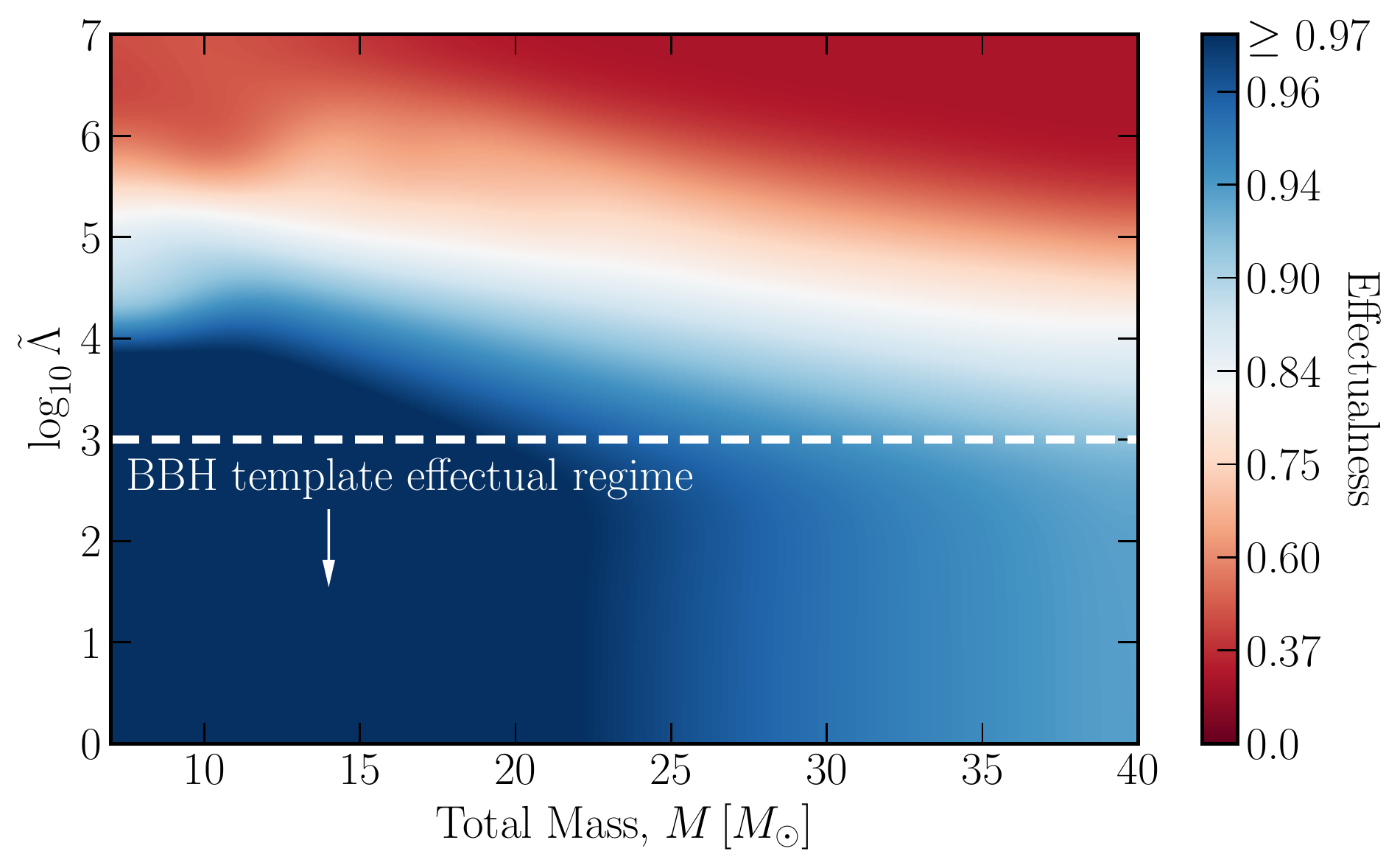}
    \caption{The ability of a standard BBH template bank to recover signals with large tidal Love numbers. In particular, we show the effectualness of a BBH template bank for the inspiral-only signal from binaries with large Love numbers as a function of binary total mass $M$ and effective tidal parameter $\tilde{\Lambda}$; see (\ref{eqn:lambdatilde}). We find that the effectualness drops significantly for $\tilde{\Lambda}\gtrsim10^3$, meaning that searches conducted with BBH template banks are much less sensitive to putative signals in this region of the parameter space. In this paper, we therefore perform a search using a template bank which includes the effect of $\tilde{\Lambda}$. Note that we allow for spins in the BBH template model --- this means that the observed reduction in effectualness due to $\tilde{\Lambda}$ cannot be recouped by allowing for spins. }
    \label{fig:effectualness}
\end{figure}

We first assess the ability of a BBH search to recover signals with large tidal Love numbers by computing the effectualness~\cite{Damour:1997ub, Buonanno:2009zt}, $\varepsilon$, of a BBH waveform model to such signals. The effectualness is defined as
\beq
\varepsilon\left(h_{\mathrm{tidal}}\right) \equiv \max _{t_c, \phi_c,\boldsymbol{p}_{\rm b b h}}\left[h_{\mathrm{tidal}} \mid h\left(\boldsymbol{p}_{\rm b b h}\right)\right]\,,
\label{eqn:effectualness}
\eeq
where the square bracket denotes the normalized inner product, $h_{\mathrm{tidal}}$ is the inspiral-only tidal waveform considered in this paper, $h\left(\boldsymbol{p}_{\rm b b h}\right)$ is the full inspiral-merger-ringdown (IMR) BBH waveform \texttt{IMRPhenomD}~\cite{Khan:2015jqa}, $t_c$ and $\phi_c$ are the phase and time of coalescence respectively, and $\boldsymbol{p}_{\rm b b h}$ represents the BBH masses and aligned-spins. The effectualness (\ref{eqn:effectualness}) can therefore be interpreted
 as the fraction of SNR of the tidal waveform retained by using the best-matched BBH template. The effectualness is a particularly important measure of sensitivity  as the resulting fractional sensitive volume achieved relative to the optimal sensitivity scales as $\varepsilon^3$. To compute \eqref{eqn:effectualness}, we follow the same procedure as in Ref.~\cite{Chia:2022rwc}, whereby we use the differential evolution algorithm to maximize over the intrinsic parameters $\boldsymbol{p}_{\rm b b h}$ and Fourier transforms to maximize over $t_c$ and $\phi_c$.
 
\vskip 4pt

In Fig.~\ref{fig:effectualness} we demonstrate the loss in effectualness of a BBH waveform model to signals with large Love numbers as a function of the binary total mass, $M$, and the mass-weighted tidal parameter, $\tilde{\Lambda}$, which depends on the individual component Love numbers; see (\ref{eqn:lambdatilde}) below.
We compute the effectualness for total masses $7\,\mathrm{M}_\odot < M < 40\,\mathrm{M}_\odot$, as this reflects the range that we will consider in our search. We find that for signals with $\tilde{\Lambda} \lesssim 10^3$, a high effectualness of $\varepsilon > 0.97$ is achieved (this is equivalent to retaining $\varepsilon^3 > 0.9$ of sensitive volume, which is the typical standard used in template bank construction). Since the $\tilde{\Lambda} \lesssim 10^3$ range encompasses both BBHs and solar-mass BNS systems, which have $\Lambda=0$~\cite{Chia:2020yla, Charalambous:2021mea, Kol:2011vg, Binnington:2009bb, Damour:2009vw} and $\Lambda \lesssim 10^3$~\cite{TheLIGOScientific:2017qsa, Abbott:2018wiz} respectively, BBH templates are effective at detecting both of these types of binary systems.
Note that there is a mild but noticeable loss in effectualness for $M\gtrsim 25\,\mathrm{M}_\odot$, which is attributed to the additional nonlinear effects near merger that are captured in \texttt{IMRPhenomD} but not in $h_{\mathrm{tidal}}$.
Nevertheless, throughout the mass range considered, signals with $\tilde{\Lambda}\gtrsim 10^3$ would likely be missed by BBH template banks.
Since objects which are less compact than black holes and neutron stars naturally fall within the $\tilde{\Lambda}\gtrsim 10^3$ parameter space, Fig.~\ref{fig:effectualness} raises the intriguing possibility that we might have missed a wider class of new signals in the LIGO-Virgo data simply because we have not been using the right waveforms to detect them. Although model-independent burst search pipelines exist~\cite{klimenko2004, Cornish:2014kda, Klimenko:2015ypf}, they are ineffective at detecting long-duration signals which contain many orbits in band.

\vskip 4pt

It is tempting to suspect, given our discussion above, that one can probe objects with arbitrarily large values of $\Lambda$ with the LIGO and Virgo observatories. However, astrophysical bodies with $r / m \gg 10^2$ would merge at frequencies well below those measured by ground-based detectors, limiting the range of $\Lambda$ that LIGO and Virgo could probe. Using Kepler's law, one can obtain an approximate upper bound on $\Lambda$ by comparing \textit{i)} the GW frequency emitted when the binary constituents touch, and \textit{ii)} the lower bound on the detector's sensitive band, $f_{\rm low}$. Assuming the object has $k \sim \mathcal{O}(1)$ and its binary counterpart is a black hole with the same mass, the LIGO and Virgo detectors can probe
\begin{equation}
\frac{r}{m} \lesssim 270 \left( \frac{20 \text{ Hz}}{f_{\rm low}} \right)^{2/3} \left(\frac{ M_\odot}{m} \right)^{2/3} \, , \quad \Lambda  \lesssim  10^{12} \left( \frac{20 \text{ Hz}}{f_{\rm low}} \right)^{10/3}  \left(\frac{ M_\odot}{m} \right)^{10/3}  \, , \label{eqn:lambda_bound}
\end{equation}
where the bound on $\Lambda$ is obtained using (\ref{eqn:tides}). This estimate implies that ground-based detectors are capable of probing objects with $\Lambda \lesssim 10^{12}$, which is still many orders of magnitude larger than that of black holes and neutron stars.

\subsection{Overview and Summary}

Motivated by Fig.~\ref{fig:effectualness}, in this paper we expand the set of template waveforms to the $\tilde{\Lambda} \gtrsim 10^3$ region and search for signals from sources in this unexplored parameter space. Our waveform model only includes the inspiral portion of the binary coalescence because this is the only regime where the Love number imprint is analytic. Such analytic control allows us to perform a source-agnostic search without specifying the type of compact object we are looking for.

\vskip 4pt
We conducted the search over all LIGO Hanford and Livingston data that were collected in the first, second, and third observing runs (O1--O3). We found no statistically significant binary events with large Love numbers. In particular, our loudest event is triggered in O2 and has an inverse false alarm rate (IFAR) that falls within the $\sim 2 \sigma$ range of Poisson noise. This null detection places a constraint on the merger rates of such binary systems. Although we take a source-agnostic approach to the search,
we map these constraints to several models of BSM compact objects proposed in the literature.

\vskip 4pt

Despite using inspiral-only template waveforms, we are able to recover many of the known BBH signals that were reported in the searches using full IMR waveforms~\cite{LIGOScientific:2018mvr, LIGOScientific:2020ibl, LIGOScientific:2021usb, LIGOScientific:2021djp, Venumadhav:2019tad, Venumadhav:2019lyq, Olsen:2022pin, Ajit-O3b, Nitz:2018imz, Nitz:2019hdf, Nitz:2021uxj, Nitz:2021zwj}. This is also despite the fact that our templates have zero spins. 
In addition, we use the tidal waveform to perform parameter estimation on these BBH events to explore any potential biases and degeneracies in the inferred intrinsic parameters. Using these results we are able to constrain the Love number of black holes. 

\vskip 4pt
 
This work is the first-ever dedicated GW matched-filtering search for compact objects that are not only black holes or neutron stars. Our results demonstrate that analytic inspiral waveform models are well suited for these kinds of new physics searches. While we focus on the tidal deformability as a probe of new physics, accounting for different finite size effects~\cite{Thorne:1980ru, Poisson:1997ha, poisson_will_2014} and other physical effects in future searches will probe even more of this previously unexplored parameter space. Future searches~\cite{Coogan:2022qxs, Chia:2022rwc} can therefore build on this work and extend the search space 
in order to realize more of GW astronomy's vast potential for discovery.

\pagebreak

\paragraph{Outline} This paper is organized as follows: in Section~\ref{sec:love_search}, we describe some technical aspects of the methods used in this work.
In Section~\ref{sec:results}, we present the results of the matched-filtering search and discuss several observations on the parameter estimation conducted with the tidal waveform model. In Section~\ref{sec:astro_implications}, we describe some consequences of our work for black holes and neutron stars. In Section~\ref{sec:bsm_implications}, we explore the implication of our null detection on BSM physics. Here we not only discuss model-independent
constraints but also examine several models of BSM compact objects to which our constraints apply. Finally, our conclusions and outlook are presented in Section~\ref{sec:conclusions}.

\vskip 4pt

\paragraph{Notations and Convention} We refer to hypothetical objects with large Love numbers as exotic compact objects or BSM compact objects interchangably. The mass and Love number of each binary component are denoted by $m_i$ and $\Lambda_i$, where $i = 1, 2$ is the component label. We use the convention $m_1 \geq m_2$. The total mass is $M = m_1 + m_2$ and $\eta = (m_1 m_2)/(m_1+m_2)^2$ is the symmetric mass ratio. We differentiate source-frame from detector-frame quantities with the superscript $\phantom{h}^{\text{src}}$. For example, the source-frame and detector-frame chirp masses are denoted by $\mathcal{M}^{\rm src}$ and $\mathcal{M}$, respectively. We work in natural units, $G=c=\hbar=1$.

\pagebreak

\section{Love Number Search} \label{sec:love_search}

In this section we present some of the technical aspects of this work. We use a modified version of the IAS matched-filter pipeline developed in Refs.~\cite{Venumadhav:2019tad, Zackay:2019kkv, Roulet:2019hzy, Olsen:2022pin, Jay-bank} (for full details, the reader should refer to the original work). In this section, we instead focus on the new developments made for the tidal waveform (\S\ref{sec:waveform}) and outline the details of the new template bank that is built for this search (\S\ref{sec:templatebank}). 

\subsection{Post-Newtonian Tidal Waveform} \label{sec:waveform}

Absent a specific compact object in mind, in \S\ref{sec:waveform_model} we focus on the inspiral stage of the binary coalescence where the physics is clean and analytically understood, with the putative new physics parameterized by the tidal parameter. 
We ignore the merger part of the waveform as that regime of a binary coalescence is model-dependent and sensitive to the sources' equations of state, thereby generally requiring inputs from numerical simulations. In \S\ref{sec:cutoff}, we describe a procedure to cutoff the near-merger part of the inspiral waveform. Care must be taken when implementing this cutoff in order to ensure that no undesirable artifacts contaminate the time-domain waveform. 

\subsubsection{TaylorF2 with Tides} \label{sec:waveform_model}

We consider a non-spinning inspiral-only waveform that is derived from the PN expansion of the relativistic dynamics of binary systems. For data-analysis purposes, the phase of the waveform must be accurate to high PN order. This requirement comes from the fact that the overlap between a potential signal in the data and the template waveform is very sensitive to phase coherence. We therefore consider point-particle phase contributions that are accurate up to 3.5 PN order (i.e., the TaylorF2 approximant \cite{Arun:2004hn, Buonanno:2009zt}), while for the finite-size contributions we consider the leading-order tidal effect, which appears at 5PN \cite{nrgr, Flanagan:2007ix}. Specifically, the phase of our waveform is
\beq
\begin{aligned}
    \psi_{\textrm{ins}}(f) & = \psi_{\mathrm{TF2}}(f) + \psi_{\mathrm{Love}}(f) \\ & = 2 \pi f t_c + \phi_c - \frac{\pi}{4} + \frac{3}{128 \hskip 1pt \eta  \hskip 1pt  v^5} \left[ 1 + \left( \frac{3715}{756} + \frac{55}{9} \eta \right) v^2 + \cdots \mathcal{O}(v^7) - \frac{39 \hskip 1pt \tilde{\Lambda}}{2} v^{10}\right] \, , \label{eqn:phase}
\end{aligned}
\eeq
where $v =  (\pi M f)^{1/3}$ is the PN expansion parameter for a quasi-circular orbit and $f$ is the waveform frequency.
In the above equation we have introduced the mass-weighted Love parameter~\cite{Flanagan:2007ix,Favata:2013rwa},
\beq
\label{eqn:lambdatilde}
    \tilde{\Lambda} = \frac{16}{13}\frac{(m_1 + 12 \hskip 1pt  m_2) \hskip 1pt m_1^4 \hskip 1pt  \Lambda_1}{M^5} + 1\leftrightarrow2 \,,
\eeq
where $\Lambda_1, \Lambda_2$ are the dimensionless Love numbers of the binary constituents, cf. \eqref{eqn:tides}. 
The parameter $\tilde \Lambda$ is the combination of $\Lambda_i$ that arises at the earliest PN order, and so it is the combination most precisely measured from the data~\cite{Flanagan:2007ix}.
Meanwhile the leading-order phase in (\ref{eqn:phase}) is proportional to the chirp mass,  $(\eta v^5 )^{-1} \propto (\mathcal{M} f)^{-5/3}$, while the next-to-leading-order term is dependent on the mass ratio. The higher-order point-particle corrections can be found in e.g. Refs.~\cite{Arun:2004hn, Buonanno:2009zt}.

\vskip 4pt

Notice that the sign of the tidal term in (\ref{eqn:phase}) is negative, i.e. opposite to that of the leading phase term. Intuitively, this occurs because part of the binding energy of the binary is transferred to distorting the binary constituents, leading to a shallower overall binding potential. As we shall see in \S\ref{sec:cutoff}, for larger values of $\tilde{\Lambda}$, this has the effect of shifting the minimum of the binding energy towards larger orbital separations and hence lower orbital frequencies. Since the binding energy minimum formally defines the innermost stable orbit of the binary system, one must implement an appropriate cutoff before the inspiral waveform ceases to be valid.

\vskip 4pt

For the amplitude, we use the leading-order PN contribution of the inspiral:
\beq
    A_{\mathrm{ins}}(f) = \pi^{2 / 3} \sqrt{\frac{5} {24}}\frac{ \mathcal{M}^{5 / 6}}{D}f^{-7/6} \, , \label{eqn:amplitude}
\eeq
where $D$ is the luminosity distance to the source. Since higher-order corrections to $A_{\mathrm{ins}}$ do not substantially affect matched-filtering, they are neglected in this work for simplicity.

\vskip 4pt

Equations~(\ref{eqn:phase}) and (\ref{eqn:amplitude}) imply that there is effectively a three dimensional intrinsic parameter space spanned by $\{\mathcal{M}, \eta, \tilde{\Lambda} \}$. These are the parameters over which we will perform our search. Notice that in order to reduce the dimensionality of parameter space, and therefore the size of our template bank, we have set the spins to zero. Spins already start contributing at 1.5PN order and their effect on the phasing is well known to be partially degenerate with that of $\eta$, which first appears at 1PN order~\cite{Cutler:1994ys, Poisson:1995ef, Baird:2012cu}. It is therefore conceivable that spins could be partially degenerate with the large tidal 5PN contribution, which we shall investigate in \S\ref{sec:lambda-spin-degeneracy}. 
Still, it is clear that for sufficiently large $\tilde{\Lambda}$ the spin effects should no longer be able to mimic the phase evolution accurately. This is shown in Fig.~\ref{fig:effectualness}, where we fixed the injected signal to have zeros spins and a range of $\tilde{\Lambda}$ but allowed for aligned spins up to $|\chi_{1,2}| <0.99$ when maximising over BBH parameters in our effectualness computation. It is clear that a standard BBH bank would have missed putative signals with $\tilde{\Lambda} \gtrsim 10^3$. Future searches should therefore look to additionally include spin effects, such as those from the spin-induced quadrupole~\cite{Chia:2022rwc,Chia:2020psj}.

\subsubsection{Smooth Waveform Truncation} \label{sec:cutoff}

\begin{figure}[t!]
    \centering
    \includegraphics[width=0.8\textwidth]{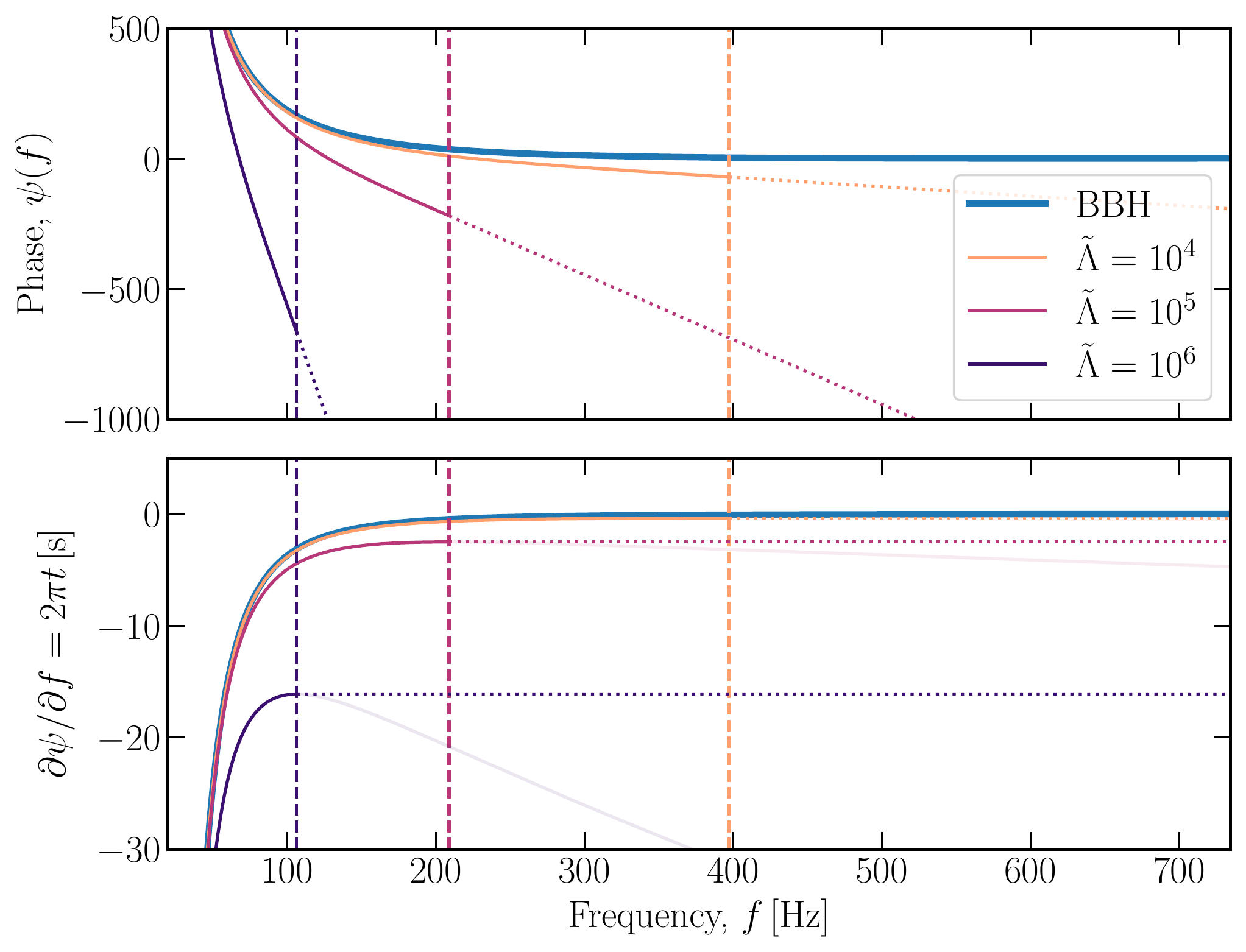}
    \caption{ The phase model of the waveforms used for our template bank and parameter estimation. We use an inspiral-only PN model for the phases including the contribution from the effective tidal parameter $\tilde{\Lambda}$;~see (\ref{eqn:phase}). We implement a frequency cutoff for our waveforms at $f_c$ (shown by vertical lines) where the phase derivative is zero, corresponding to the end of the validity of the stationary phase approximation.
    \textit{(top)} Waveform phases $\psi$ for a range of  $\tilde{\Lambda}$, with masses fixed to $m_1 = m_2 = 3 M_\odot$. We use the derivative of the phase at $f_c$ to extrapolate beyond $f_c$ to enforce $C^1$ continuity. \textit{(bottom)} Same as the top panel, except, we show the derivative of $\psi$, which is proportional to the instantaneous time. For reference, we show the
    naive extrapolation of $\psi_{\rm inspiral}$ from (\ref{eqn:phase}) beyond $f_c$ with reduced opacity, indicating that the stationary phase approximation breaks down in this regime as a moment in time would correspond to two frequency solutions.
    }
    \label{fig:phase_derivative}
\end{figure}

An important property of the waveform model in \S\ref{sec:waveform_model} is that for large values of $\tilde{\Lambda}$ the derivative of the phase can reach a maximum earlier than the typical frequency cutoff for the TaylorF2 model,
signaling the end of the inspiral region (a similar observation was first pointed out in Ref.~\cite{Chia:2022rwc} for spin-induced quadrupole moments). This is evident from \eqref{eqn:phase}, where the tidal term being negative causes the derivative to reach a maximum, cf. Fig.~\ref{fig:phase_derivative}. 
At higher frequencies, the adiabatic approximation that holds during inspiral
is no longer a good physical representation of the binary dynamics beyond that stage. For a BBH, $\tilde{\Lambda} = 0$ and therefore this maximum in the TaylorF2 phase is absent.
For $\tilde{\Lambda} \sim 10-10^3$ close to BNS values, this maximum is normally well above the innermost stable circular orbit (ISCO) frequency and is naturally cutoff when constructing a full IMR waveform. However, for systems with large values of $\tilde{\Lambda}$, this maximum can occur well below the ISCO frequency. See Fig.~\ref{fig:phase_derivative} for the locations of the maxima for various values of $\tilde{\Lambda}$.

\vskip 4pt

From a practical perspective, truncating the waveform at the peak of the phase derivative also ensures that the corresponding time domain waveform is monotonically increasing in frequency. This can be understood from the fact that by definition of a frequency-dependent phase, $\partial \psi/\partial f = 2\pi t$. Without a truncation, the time domain waveform at any particular time would have contributions from two different frequencies, signaling a breakdown of the stationary phase approximation~\cite{Thorne1980Lectures, Cutler:1994ys, Droz:1999qx} assumed in the derivation of (\ref{eqn:phase}). This behaviour is shown in the bottom panel of Fig.~\ref{fig:phase_derivative}, where an instantaneous moment in $t$ would correspond to two different frequency solutions had the cutoff not been introduced. To prevent this behaviour, we introduce the cutoff 
\begin{equation}
    f_c = \text{argmax}_f \hskip 4pt \left[ \frac{\partial \psi_{\mathrm{ins}}(f)}{\partial f } \right] \, , \label{eqn:fcutoff}
\end{equation}
which can be computed analytically from (\ref{eqn:phase}), and truncate the waveforms at the frequencies beyond these maxima. 

\vskip 4pt

It is instructive to compare $f_c$ with the GW frequency at which the binary components touch. We obtain an approximate analytic expression for $f_c$ by focusing on the 0PN and 5PN tidal terms (\ref{eqn:phase}) and setting the second derivative of the phase to zero, after which we find
\begin{equation}
    f_c \approx \frac{2^{9/10}}{39^{3/10} M \pi \tilde{\Lambda}^{3/10}} = 126 \text{Hz} \left( \frac{20 M_\odot}{M} \right) \left( \frac{10^4}{\tilde{\Lambda}} \right)^{3/10} \, . \label{eqn:cutoff_analytics}
\end{equation}
On the other hand, the GW frequency emitted when the binary components touch can be estimated via Kepler's third law, where we equate the sum of the bodies' radii $r_{1,2}$ with the binary separation:
\begin{equation}
    f_{\rm touch} = \frac{1}{\pi} \sqrt{\frac{m_1 + m_2}{(r_1 + r_2)^3} } \, . \label{eqn:touch_analytics}
\end{equation}
For simplicity, we assume one of the components is a black hole (approximated by $r_1 \gg r_2$) and both components have the same mass. Using (\ref{eqn:lambdatilde}) and the definition for the component Love number (\ref{eqn:tides}), we obtain
\begin{equation}
    f_{\rm touch} \approx \frac{2^{3/2} k^{3/10}}{3^{3/10} M \pi \tilde{\Lambda}^{3/10}} = 413 \text{Hz} \hskip 2pt k^{3/10}  \left( \frac{20 M_\odot}{M} \right) \left( \frac{10^4}{\tilde{\Lambda}} \right)^{3/10} \, . \label{eqn:touch_analytics_simple}
\end{equation}
This is the same derivation which led to the upper bound (\ref{eqn:lambda_bound}). Comparing (\ref{eqn:cutoff_analytics}) and (\ref{eqn:touch_analytics_simple}) we find $f_c / f_{\rm touch} \approx 0.3 \hskip 1pt k^{-3/10}$ --- the inspiral waveform is truncated before the objects touch, as it should if our waveform is intended to only represent the inspiral regime of the binary system. 

\vskip 4pt

The cutoff $f_c$ introduces a sharp discontinuity in both the phase and the amplitude, which has a variety of consequences throughout the search pipeline. For example, when constructing the template bank, we perform a singular value decomposition (SVD) on the phase. Having discontinuities in the phase and phase derivatives leads to the spurious result that a large number of basis functions is required to accurately model the phase evolution. 
Additionally, we require the time domain waveform~\cite{Venumadhav:2019tad} at various points during the matched-filter search, and a sharp discontinuity in the amplitude in the frequency domain leads to undesirable features in the time-domain waveform. 
In particular, multiplying the Fourier-domain waveform by a step function corresponds to convolving the time-domain waveform with a sinc function, which leads to a ringing artifact in the time-domain waveform after the cutoff due to the Gibbs phenomenon.
For these reasons we need to enforce continuity in both the phase and amplitude separately.

\vskip 4pt

For the phase, we enforce $C^{1}$ continuity by appending an overall constant and linear-in-$f$ contribution to the phase beyond the cutoff.
As we shall elaborate in \S\ref{sec:templatebank}, the $C^{1}$ continuity significantly reduces the number of dimensions needed in the SVD of the phase in order to achieve high effectualness for the template bank. Our full phase function is given by:
\beq
\psi(f) = \psi_{\textrm{ins}}(f) \hskip 1pt \theta(f_c-f) + \left[ \alpha_0+ \alpha_1 f\right] \hskip 1pt  \theta(f-f_c)\,, \label{eqn:total_phase}
\eeq
where we solve for the $\alpha$'s by enforcing $C^{1}$ continuity at the boundary. See the top panel of Fig.~\ref{fig:phase_derivative} for an illustration of $\psi$ for a few representative values of $\tilde{\Lambda}$.

\vskip 4pt

Meanwhile, the location of the sharp amplitude cutoff in the frequency domain varies with the model parameters.
This makes it difficult to alleviate ringing by using a common windowing function across our model space. Employing a window function to smoothly reduce the waveform to zero at the amplitude cutoff sacrifices part of the inspiral within the sensitive band of the detectors.
This can reduce the SNR extracted by up to 20\%, which is undesirable.

\vskip 4pt

We therefore aim to stitch on a \textit{smooth cutoff} to exponentially damp the artificial ringing. 
In particular, we only make it approximately $C^{1}$ continuous at the boundary and damp the amplitude very quickly using a sigmoid function. The amplitude after the cutoff is given by
\beq
A_{\mathrm{c}}(f) = A_{\mathrm{ins}}(f_c)\left(\frac{f_c}{f}\right)^{7/6}\left[1 - \frac{1} { 1 + e^{-(f - \beta f_c)}}\right]\,,
\eeq
where $A_{\mathrm{ins}}(f_c)$ is the amplitude of the inspiral (\ref{eqn:amplitude}) evaluated at the cutoff and $\beta$ is a constant that controls the rate at which the amplitude is exponentially damped. We find that using $\beta=1.2$ works well to preserve continuity at the boundary, while not introducing artifacts into time domain waveforms. The final amplitude is therefore given by
\beq
A(f) = A_{\mathrm{ins}} \hskip 1pt \theta(f_c-f) + A_{\mathrm{c}}(f)  \hskip 1pt \theta(f-f_c)\,. \label{eqn:total_amplitude}
\eeq
An illustration of this cutoff and how it affects the amplitude can be seen in Fig.~\ref{fig:Aref_effectualness}. 

\vskip 4pt 

To summarize, the Fourier-domain waveform used throughout the rest of the paper is
\beq
h_{\rm tidal}(f) = A(f) \hskip 1pt e^{-i\psi(f)}\,, \label{eqn:total_wf}
\eeq
where the total phase and amplitude are found in (\ref{eqn:total_phase}) and (\ref{eqn:total_amplitude}), respectively.

\subsection{A New Template Bank} \label{sec:templatebank}

We build a new template bank based on the inspiral-only waveform described in \S\ref{sec:waveform_model}. The bank is constructed using the geometric-placement technique described in Refs.~\cite{Roulet:2019hzy, Jay-bank}. For details on the bank construction method, refer to the original works. We motivate our choice of search parameter space in \S\ref{sec:search_params} and provide additional details of the bank in \S\ref{sec:bank_details}.

\subsubsection{Search Parameter Space} \label{sec:search_params}

The parameter space region over which we perform our search is defined below. The cuts were made with the aim of maximising the space over which we search, together with computational cost considerations:

\begin{itemize}

    \item \textit{Mass lower bound ---}  The lower bound of our mass region is informed by the fact that the number of templates required to achieve a predetermined template bank efficiency grows significantly as we approach the low-mass regime (roughly as $N_\mathrm{templates}\propto \mathcal{M}_\mathrm{min}^{-8/3}$~\cite{Owen:1998dk}). For instance, in the template bank constructed in Ref.~\cite{Roulet:2019hzy}, the number of templates needed in the BBH mass range ($m_1, m_2 > 3M_\odot$) is only $\sim 1.5 \times 10^4$ while the relatively narrow BNS range ($1 M_\odot < m_1, m_2 < 3 M_\odot$) requires $\sim 6 \times 10^4$ templates. This is the case because the smaller the binary mass, the larger the number of orbital cycles seen by the LIGO and Virgo detectors. This makes it necessary to densely sample the low-mass parameter space in order to preserve phase accuracy to $\ll \mathcal{O}(1)$ radians. 

    \item \textit{Mass upper bound ---} The upper mass bound is motivated by the fact that the model-independent Coherent WaveBurst (cWB) pipeline~\cite{klimenko2004, Klimenko:2015ypf} used by the LIGO-Virgo-Kagra (LVK) Collaboration is already well suited to detect any new short duration signal (corresponding to high masses) which is not modeled by BBH templates. 
    From the GWTC-2 catalog~\cite{LIGOScientific:2020ibl}, we observe that the cWB pipeline was able to detect BBH systems with source-frame masses $M^{\rm src} \gtrsim 40 M_\odot$ and $\mathcal{M}^{\rm src} \gtrsim 15 M_\odot$ which were also detected by matched-filtering searches. Below this mass range, the cWB's detection capabilities manifestly decreases.

    \item \textit{Love number upper bound ---} While a search over as large a range as possible in $\tilde{\Lambda}$ space would have been ideal, an upper bound is inevitably imposed due to its correlation with the cutoff frequency $f_c$ in the waveform model. As shown in Fig.~\ref{fig:phase_derivative}, binary systems with larger values of $\tilde{\Lambda}$ would have lower $f_c$. We demand that a template should integrate SNR over a sufficiently wide frequency range in the data by imposing a minimum cutoff, $f_{c, {\rm min}}$, in the waveform model. That is, we require that the upper bound on the frequency grid, $f_c$, is no lower than $f_{c, {\rm min}}$ so that the frequency range entering into the pipeline computations for a given template with parameters $M, \tilde{\Lambda}$ is $f \, \in \, [\, 20, \, \text{max}( f_c , f_{c, {\rm min}} )]$ Hz.

\end{itemize}

\begin{figure}[t!]
    \centering
    \includegraphics[width=0.95\textwidth, trim=0 20 0 0]{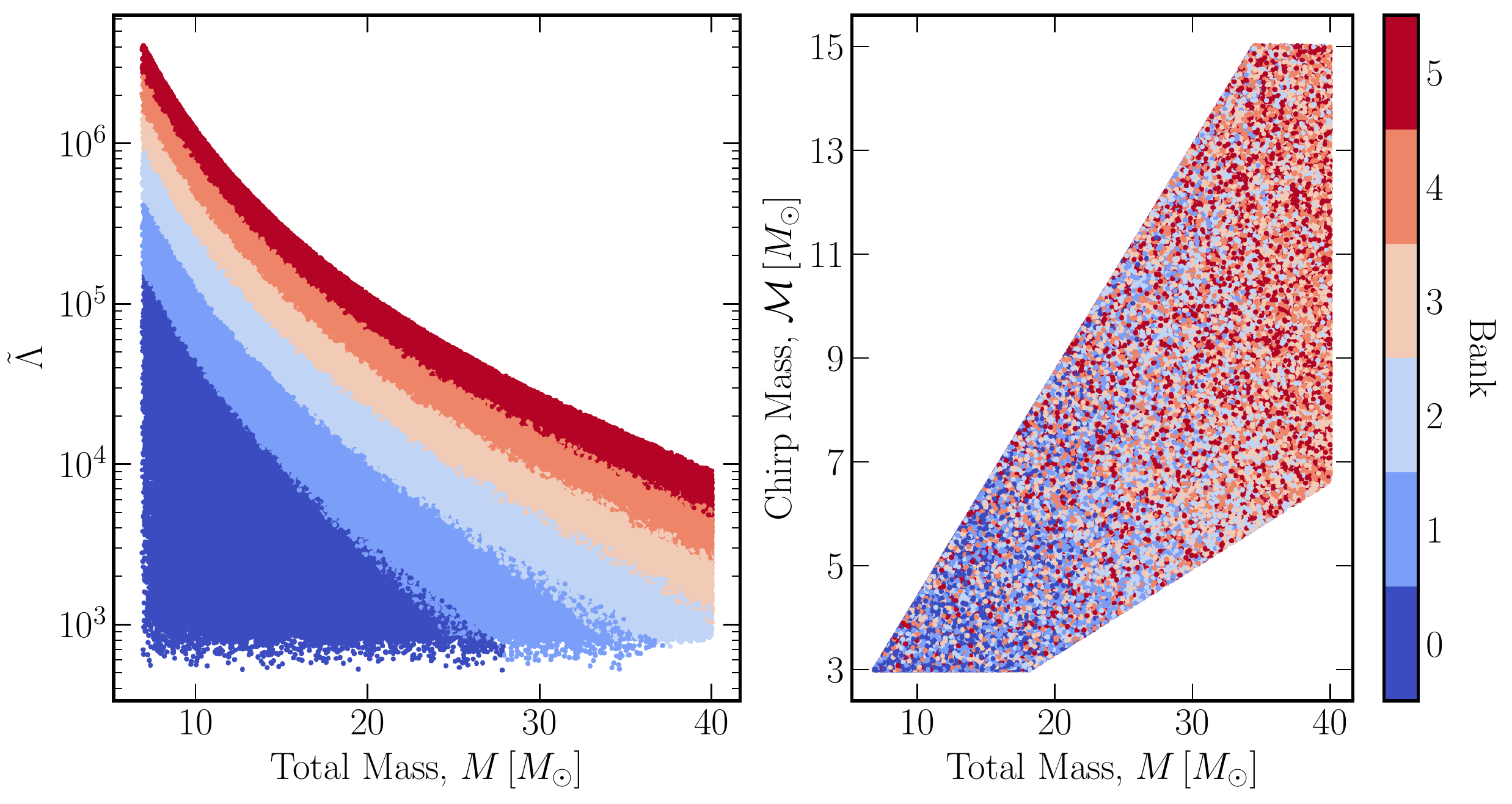}
    \caption{Parameter ranges of waveforms in our template banks. \textit{(left)}
    Our banks are divided such that waveforms in the same bank have similar normalized amplitudes $f^{7/6} A(f)$. In our case, the amplitude is essentially dependent on the reference cutoff frequencies $f_c$ (cf.~Fig.~\ref{fig:Aref_effectualness} and Table~\ref{tab:bank_details}). Hence, the lines in the $\tilde{\Lambda}-M$ plane demarcating the banks roughly correspond to lines of constant $f_c$. Considering the current frequency sensitivity of the detectors, we set the range $f_c \geq 60\text{Hz}$ for our search, which leads to the white region in the top-right being masked out.
    \textit{(right)} The detector-frame mass parameters of waveforms in the template banks.} 
    \label{fig:bank_params}
\end{figure}

\noindent Based on these considerations, we sample an initial set of points which will be used as the input for the SVD in our template bank construction below (see \S\ref{sec:bank_details}). The mass sampling has two contributions: \textit{i)} the first is sampled uniformly in the detector-frame chirp mass and $\ln q$, where $q = m_2 / m_1$ over the intervals $3 M_\odot < \mathcal{M} < 15 M_\odot, \hskip 1pt 1/18 \leq q \leq 1$; \textit{ii)} the second is uniform in the detector-frame total mass and symmetric mass ratio over the intervals $ 2 M_\odot < M < 40 M_\odot, \hskip 1pt 0.05 < \eta < 0.25  $. We then apply the parameter cuts $3 M_\odot < \mathcal{M} < 15 M_\odot$ and $M < 40 M_\odot$ to remove any samples outside this region. The final distribution of masses is shown in Fig.~\ref{fig:bank_params}.

\vskip 4pt

For the tidal Love numbers, we first sample uniformly in the log of each component Love number over the interval $\log_{10}(500) < \log_{10} \Lambda_{1,2} < 7$, and then impose a minimum cutoff frequency $f_{c, {\rm min}} = 60~\text{Hz}$. This choice of $f_{c, {\rm min}}$ is made in order to ensure that the signals are at least integrated over the frequency range $[20, 60]$ Hz and can therefore accumulate sufficient amounts of SNR in the data (see Table~\ref{tab:bank_details} for the range of $f_c$ in each bank). The lower bound in $\log_{10} \Lambda_{1,2}$ is intentionally chosen to overlap with neutron star values and avoid the $\Lambda_{1,2} = 0$ value for black holes, in order to demonstrate the ability of finite $\tilde{\Lambda}$ waveforms to detect BBHs, cf. Fig.~\ref{fig:effectualness}. The resulting distribution in $\tilde{\Lambda}$ is shown in Fig.~\ref{fig:bank_params}, where the masked region has $f_c < 60$ Hz. Despite the masking due to $f_c$, our template bank covers several orders of magnitude in $\tilde{\Lambda}$, overlapping with regions in parameter space where the effectualness of BBH template bank is significantly degraded, cf. Fig.~\ref{fig:effectualness}.

\vskip 4pt

We emphasize that we do not include spins in the template bank parameter space. This simplification is made in order to reduce the dimensionality of the waveform model, which substantially reduces the number of templates needed for an effectual bank. In Section~\ref{sec:results} we show that, despite this simplification, we are able to recover many confidently detected BBHs which overlap with our mass coverage.

\subsubsection{Bank Details and Effectualness} \label{sec:bank_details}

Having determined the search coverage above, we construct a template bank using the geometric-placement formalism described in Ref.~\cite{Roulet:2019hzy}. The key idea of this method lies in the following decomposition of the waveform phase
\begin{equation}
    \psi(f) = \overline{\psi}(f) + c_0 + c_1 f + \sum_{\alpha = 2}^{N_{\rm dim}+2} c_{\alpha} \psi_\alpha(f) \, ,
\label{eq:calpha}
\end{equation}
where $\overline{\psi}$ is an average phase chosen for convenience, $c_0$ and $c_1$ are coefficients that capture the overall constant and linear-in-time phase offset, $c_\alpha$ is a set of coefficients which only depend on the intrinsic source parameters, $\psi_\alpha$ is a set of orthonormal basis functions constructed such that their inner product defines a locally Euclidean space, and  $N_{\rm dim}$ is the size of the $\psi_{\alpha}$ set. The basis functions and $c_\alpha$'s are computed through an SVD performed on random phase samples drawn from our designated parameter space (as described above). $N_{\rm dim}$ and the distance between templates, $\Delta c_\alpha$, are suitably chosen so that the truncation in SVD strikes a good balance between the effectualness of the bank and the density of templates in $c_{\alpha}$ space (see discussion below). 

\begin{figure}[t!]
    \centering
    \includegraphics[width=\textwidth, trim=0 20 0 0]{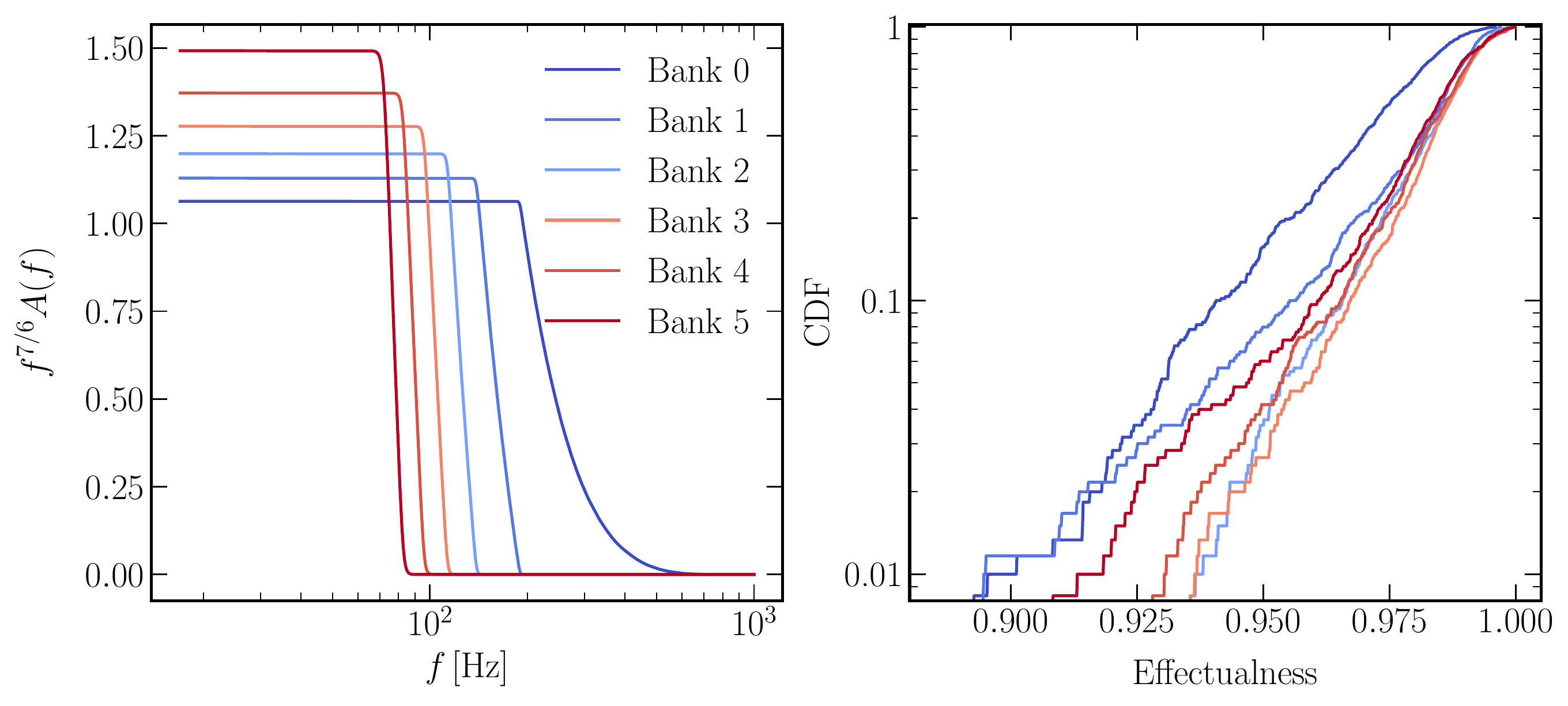}
    \caption{ \textit{(left)} We analytically model the waveform amplitudes in our template banks using (\ref{eqn:total_amplitude}). Each bank corresponds to a different (normalized) reference amplitude. Waveforms in different banks are essentially distinguished by their cutoff frequencies $f_c$. Note that the amplitude for $f>f_c$ is damped exponentially (instead of using a sharp cutoff) in order to prevent artificial ringing effects in time-domain waveforms. \textit{(right)} The cumulative distribution function of the average effectualness $\varepsilon$ of the template banks, tested on waveforms whose parameters are drawn from the parameter space spanned by the respective banks.
    We find, on average, that $\varepsilon \gtrsim 0.9$ for $99\%$ of tested waveforms for each bank. Note that each bank is further divided into three subbanks and the effectualness shown here corresponds to their average. Note that the loss of effectualness is dominated by subbank 0, which has the lowest $\mathcal{M}$ range, and therefore contains the most templates (we relax the effectualness requirement here to reduce the number of templates). 
    We have tested that excluding the lowest subbank in each bank, we get $\varepsilon > 0.96$ for $99\%$ of tested waveforms.
    }
    \label{fig:Aref_effectualness}
\end{figure}

\vskip 4pt

We divide the parameter space into banks and subbanks to ensure that differences in noise distributions between each subbank do not unfairly downweight potentially viable candidate triggers. Instead of dividing the banks based on mass ranges and the subbanks based on a set of reference amplitudes, as in the original work in Ref.~\cite{Roulet:2019hzy}, in this paper we adopt a reversed strategy from Ref.~\cite{Jay-bank}. Specifically, we partition the banks based on a set of reference amplitudes by demanding the overlap in amplitude between a sampled waveform and its closest reference waveform to be $>0.96$, and then divide each bank into subbanks with separate ranges in $\mathcal{M}$.

\begin{table}[t!]
    \def\arraystretch{1.4}
    \centering
    \begin{tabular}{|c||c||c|c|c|c|c|}
    \hline\hline
    Bank  & $f_{c}$ range [Hz] & $N_{\rm subbanks}$ & $\mathcal{M}_{\rm boundaries}$ $ [M_\odot]$ & $N_{\rm dim}$  & $\Delta c_{\alpha}$ & $N_{\rm templates}$   \\ \hline
    0 & (158,  633) & 3 & [3.0, 3.7, 4.9, 12.1] & 3 & [0.9, 0.75, 0.5] & 9811 \\ \hline
    1 & (116,  158) & 3 & [3.0, 4.6, 6.6, 15.0] & 3 & [0.9, 0.5, 0.5] & 6907 \\ \hline
    2 & (94 , 116)  & 3 & [3.0, 5.2, 7.8, 15.0] & 3 & [0.75, 0.5, 0.5] & 4979 \\ \hline
    3 & (80 , 94) & 3 & [3.0, 5.6, 8.3, 15.0] & 3 & [0.75, 0.5, 0.5] & 2833 \\ \hline
    4 & (69 , 80) & 3 & [3.0, 5.7, 8.6, 15.0] & 3 & [0.75, 0.5, 0.5] & 2068 \\ \hline
    5 & (60 , 69) & 3 & [3.0, 5.9, 8.8, 15.0] & 3 & [0.75, 0.5, 0.5] & 1615 \\ \hline
    \multicolumn{5}{|c}{} & \multicolumn{1}{c|}{Total} & 28213 \\ \hline \hline
    \end{tabular}
    \caption{Additional details of the template banks. The banks are partitioned in such a way that the overlaps between the sampled waveform amplitudes and their respective reference amplitudes are highest, which is determined by a \texttt{KMeans} algorithm; see Figs.~\ref{fig:bank_params} and~\ref{fig:Aref_effectualness}. This partitioning directly correlates with the waveform cutoff frequencies $f_c$ of the sampled waveforms, and the ranges of $f_c$ covered by the banks are shown in this table. $N_{\rm subbanks}$ is the number of subbanks in each bank; $\mathcal{M}_{\rm boundaries}$ lists the chirp mass boundaries between subbanks 0, 1, and 2 in each bank over the $\mathcal{M} \in (3, 15) M_\odot$ interval; $N_{\rm dim}$ is the number of independent $c_{\alpha}$ dimensions for the intrinsic source parameters; $\Delta c_{\alpha}$ is the list of grid spacings that we choose for subbanks [0,1,2] in each bank, which essentially determine the number of templates in each bank $(N_{\rm templates})$.}
    \label{tab:bank_details}
\end{table}

\vskip 4pt

Here, we briefly outline the procedure used for creating the template banks based on Ref.~\cite{Jay-bank}.
The banks' reference amplitudes are obtained through a \texttt{KMeans} algorithm~\cite{DBLP:journals/corr/BuitinckLBPMGNPGGLVJHV13}, which identifies centroids in the space of waveform amplitudes that maximize the average overlap between the resulting reference amplitudes and the sampled waveforms associated to each centroid (\ref{eqn:total_amplitude}). 
In order to achieve the requirement of minimum amplitude overlap $>0.96$, we find it is adequate to divide the parameter space into six banks. 
To compute this overlap we use a reference power spectral density (PSD) that is obtained by applying Welch's method~\cite{1161901} over 50 random O3 LIGO-Virgo data files and taking the $10$th percentile of the sample of PSDs in order to downweight the distortions arising from the spectral lines and loud glitches.
The normalized reference amplitudes are shown in Fig.~\ref{fig:Aref_effectualness}, where we observe how the \texttt{KMeans} algorithm essentially differentiates the reference amplitudes through their reference $f_c$. This direct correlation with reference cutoff frequencies is responsible for the stratification of template banks in $\tilde{\Lambda}-M$ space seen in Fig.~\ref{fig:bank_params}. In Table~\ref{tab:bank_details}, we list the ranges of $f_c$ spanned by each bank. Note that Fig.~\ref{fig:Aref_effectualness} also provides an instructive illustration of the smooth exponential damping of waveform amplitudes near the cutoff, as described in (\ref{eqn:total_amplitude}). Finally we divide each bank into three subbanks, with the boundaries between subbanks obtained by requiring that the number of samples of $\mathcal{M}$ in the mass region described in \S\ref{sec:search_params} to be equal for each subbank. As a result, the lower subbanks, which concentrate on low $\mathcal{M}$, tend to have shorter ranges compared to the higher subbanks. For instance, subbanks 0, 1, and 2 in Bank 0 have the following $\mathcal{M}$ ranges: $(3, 3.7)M_\odot , (3.7, 4.9)M_\odot$ and $ (4.9, 12.1) M_\odot$; see Table~\ref{tab:bank_details} for the subbank boundaries in  $\mathcal{M}$  for all banks.

\vskip 4pt

Empirically, we found that a minimum of $N_{\rm dim}=3$ was sufficient to achieve a reasonable level of effectualness. Intuitively, the first three basis coefficients $\{c_2,c_3,c_4\}$ in (\ref{eq:calpha}) correspond to the $\{\mathcal{M}, \eta, \tilde{\Lambda} \}$ parameters in the waveform model's phase; see \S\ref{sec:waveform_model}. Crucially, had we not enforced $C^1$ continuity in the phase after the cutoffs (a procedure on which we elaborated in \S\ref{sec:cutoff}), sharp features that otherwise would have been present at the phase cutoffs would have resulted in unnecessary correlations between basis functions of the SVD. In that case, a larger number of $c_\alpha$ dimensions $N_{\rm dim} \sim \mathcal{O}(10)$, and therefore a significantly larger number of templates, would be needed to achieve the same level of effectualness that our modified $C^1$ continuous waveform requires. 
Although we find that three basis functions capture nearly all of the phase information of our smoothed waveform, we can achieve additional accuracy in the templates by including information beyond the three-dimensional $c_{\alpha}$ space.
To achieve this accuracy improvement efficiently, we use ten basis $\psi_{\alpha}$ functions but we train a \texttt{RandomForestRegressor}~\cite{DBLP:journals/corr/BuitinckLBPMGNPGGLVJHV13} to use the first three $c_{\alpha}$ coefficients to predict the next seven~\cite{Jay-bank}.
We estimated this to provide an additional $\sim 1\%$ improvement in effectualness without affecting the template bank size. Note that, given the partial degeneracy between the intrinsic parameters (detailed later in \S\ref{sec:lambda-spin-degeneracy}), using two instead of three $c_\alpha$ dimensions  (and predicting the rest of the coefficients with \texttt{RandomForestRegressor}) might possibly be enough. This could, in principle, reduce the number of templates in our banks, and we explore this direction in future work.

\vskip 4pt

For each subbank, we aim to achieve an effectualness of $\varepsilon >0.96$ for 99\% of the waveforms used in the effectualness test by tuning $\Delta c_\alpha$ individually. We found that subbanks 1 and 2 easily achieve this criterion for all banks with a modest number of templates, $\mathcal{O}(10^2-10^3)$. 
On the other hand, each subbank 0 would require a rather exorbitant number of templates $\sim \mathcal{O}(10^4-10^5)$ to achieve a similar level of effectualness. To minimize computational cost, for all subbank $0$'s we compromise by only requiring $\varepsilon >0.90$ for 99\% of the tested waveforms, leading to a substantial reduction in the number of templates, $\mathcal{O}(10^3)$. Since the sizes of subbank $0$'s still dominate over those of the higher subbanks, the average effectualness of $\varepsilon >0.90$ for 99\%, shown in Fig.~\ref{fig:Aref_effectualness}, can be interpreted as a conservative estimate of the overall banks' effectualness. Table~\ref{tab:bank_details} summarizes our choices for $\Delta c_\alpha$ and the total number of templates used in this work.

\pagebreak

\section{Observations and Inferences } \label{sec:results}

In this section we present the results of the matched-filtering search (\S\ref{sec:observations_pe}) and the parameter estimation (\S\ref{sec:pe}) using the tidal waveform developed in Section~\ref{sec:love_search}.

\subsection{Matched-Filtering Results} \label{sec:observations_pe}

After constructing the template banks as described in Section~\ref{sec:love_search}, we use the IAS pipeline developed in Refs.~\cite{Venumadhav:2019tad, Zackay:2019kkv, Roulet:2019hzy, Olsen:2022pin} for conducting a matched-filtering search on all the Hanford and Livingston detector data from the O1-O3 observing runs. The details of the IAS pipeline are summarized in Ref.~\cite{Venumadhav:2019tad} but we present a brief overview below.

\vskip 4pt

The pipeline first preprocesses the strain data. This includes three steps: First, we measure the PSDs via the Welch method~\cite{1161901}. Second, we create ``holes" in the data to excise bad data segments such as abrupt transients (``glitches") or excess power localized to particular bands and timescales \cite{Venumadhav:2019tad}. Third, we fill the created holes using an ``inpainting filter" (cf. Fig.~6 of \cite{psd_drift}). The pipeline then performs matched-filtering separately for the Hanford and Livingston detectors, collecting triggers above a certain SNR threshold. Note that the SNR calculation takes into account leading-order non-stationarity in the data (also called the PSD drift correction \cite{psd_drift}). The pipeline also checks whether the SNR of each trigger builds up the right way with frequency and triggers that fail any of these ``split tests" are vetoed. A coincidence analysis is then conducted over the remaining triggers in order to ensure that physical signals are not separated by more than the detectors' light crossing time of $\sim 10$ ms and that phase differences are consistent with the measured time differences for triggers with the same intrinsic parameters in both detectors. Subsequently, a ranking statistic is used to downweight heavy tails in the trigger distribution caused by loud transient glitches (which survive all veto tests) in each detector \cite{Venumadhav:2019lyq,Venumadhav:2019tad}. A multi-detector statistic referred to as the ``coherent score", derived from the relative amplitude, phase and detector arrival times, is used to further improve the ranking score \cite{Olsen:2022pin, Nit17}. Finally, the pipeline repeats the full analysis 2000 times on unphysical time slides of the multi-detector data (i.e., artificially shifting one of the two detectors by more than the light crossing time between them and finding spurious ``coincident" triggers) in order to estimate the background noise, which allow us to compute the false alarm rates of our coincident trigger list.

\vskip 4pt

We discuss the final list of triggers with large Love numbers in \S\ref{sec:love_triggers} and also show that we are able to recover many previously detected BBH mergers in \S\ref{sec:known_bbhs}.

\subsubsection{Marginal Triggers of Love} \label{sec:love_triggers}

In Table~\ref{tab:signalsFound_new}, we present the details of the top few candidates with large Love numbers, including the event datetime labels, GPS times, the banks in which they were triggered, the best fit templates' detector-frame chirp mass $\mathcal{M}$ and leading 5PN tidal parameter $\tilde{\Lambda}$, the squared SNR detected by the Hanford and Livingston detectors, $\rho_{\rm H}^2$ and $\rho_{\rm L}^2$, and the inverse false alarm rate (IFAR) in units of years per bank.\footnote{The IFARs are computed within each bank and are given in terms of years based on total analysis times of 46, 118, 106 and 96 days for O1, O2, O3a and O3b Hanford--Livingston coincidences. The IFARs are reported for each bank as we do not impose a prior over banks. To obtain an approximate estimate of the IFAR across all banks one can divide the reported IFAR by a trials factor of $N_{\rm trial}=6$ (number of banks used in this work). \label{footnote:ifar}} We only list candidate events with IFAR $> 1$ year per bank.

\begin{figure}[t!]
    \centering
    \includegraphics[width=0.6\textwidth]{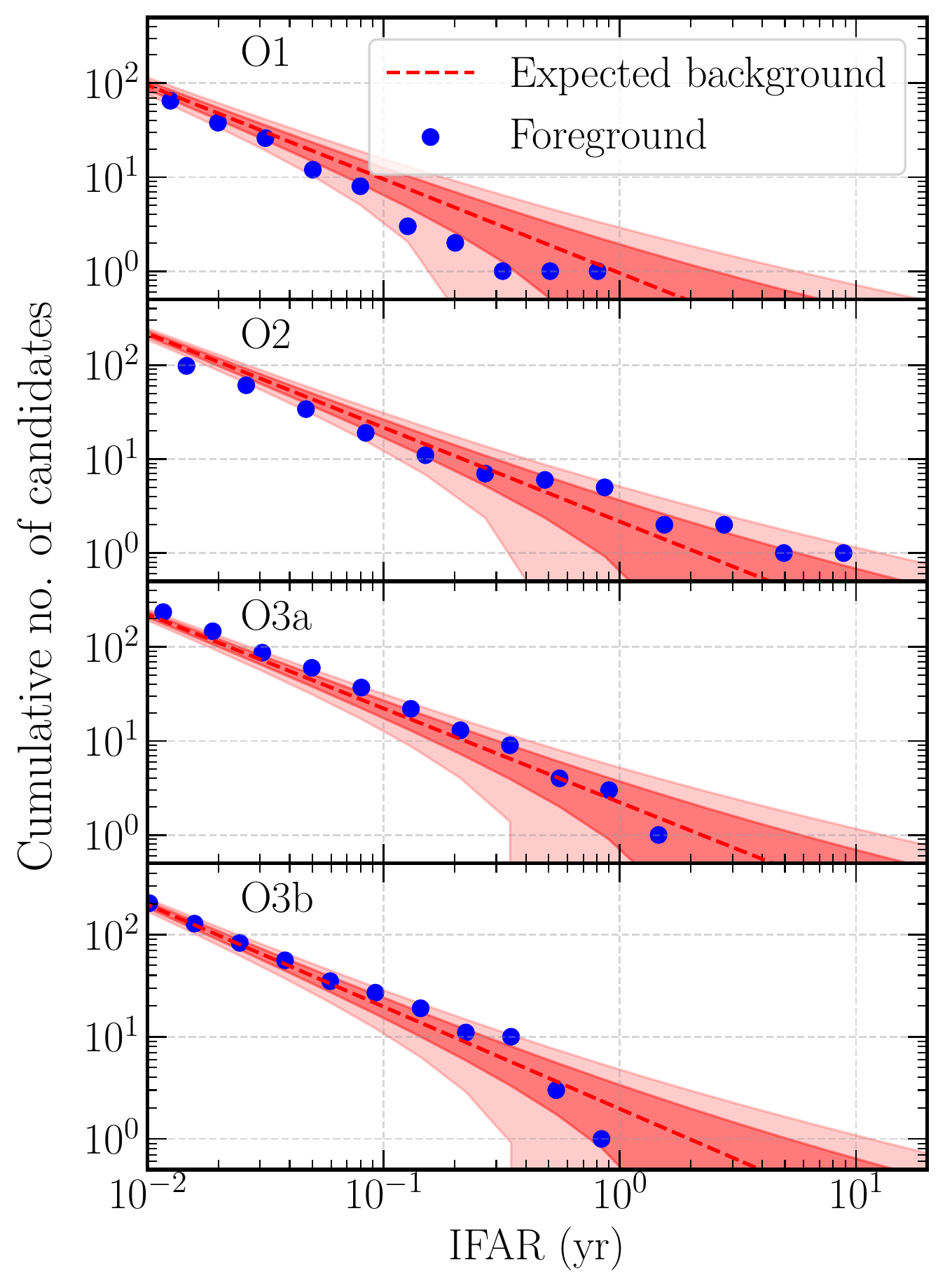}
    \caption{Comparing the distribution of the candidates in our search to the expected background rate separately for the different LIGO-Virgo observing runs. Excluding the events already reported in previous searches, we show the cumulative number of new candidates found in our search (labeled as foreground) as a function of their inverse false alarm rate (IFAR). The background rate is estimated by redoing our coincidence analysis on 2000 runs of unphysical timeslides (i.e., artificially shifting the detectors by more than the light crossing time between them and then finding the rate of spurious coincidences), and the shaded regions correspond to the $1\sigma$ and $2\sigma$ Poisson background. We see that the distribution of our candidates is fairly consistent with the background.  }
    \label{fig:ifar}
\end{figure}

\vskip 8pt

\bgroup
\begin{table}[h!]
    \centering
    \setlength\tabcolsep{4pt}
    {\tabulinesep=1.0mm
    \begin{tabular}{|c|c|c|cc|cc|c|}
        \hline
        Marginal Trigger Datetime &  GPS Time & Bank & $\mathcal{M}$ [$M_\odot$] & $\tilde{\Lambda}$ &  $\rho_{\rm H}^2$ & $\rho_{\rm L}^2$ & IFAR$^{\ref{footnote:ifar}}$ [yr]  \\
         \hline
\rowcolor{gray!30} 170212\_050900 (O2) & 1170911358.28 & \texttt{4} & 3.8 & 557386 & 34.8  & 42.1 & 8.86  \\
170317\_141415 (O2) & 1173795273.38 & \texttt{4} & 3.3 & 368151 & 47.8  & 27.9 & 3.35  \\
\rowcolor{gray!30} 190524\_123941 (O3a) & 1242736799.90 & \texttt{5} & 4.0 & 73984  & 48.6 & 28.8  & 1.46   \\
         \hline
     \end{tabular}}
    \caption{Our top triggers with large Love numbers and IFAR$ > 1$ year per bank, sorted by descending order in IFAR. Assuming Poisson statistics for the O2 background triggers, the top event with an IFAR of 8.86 year per bank corresponds to a p-value of $\sim20\%$ (see main text), which is not statistically significant.
    }
    \label{tab:signalsFound_new}
\end{table}
\egroup

Our two most significant candidates were triggered by Bank 4 in the O2 run. The top event has an IFAR of 8.86 years per bank -- absent a prior over banks, we use the trials factor of $N_{\rm trial} = 6$ to obtain an estimate of the average IFAR of $1.48$ years over all banks for the most significant trigger. Assuming a Poisson distribution for the background triggers, the coincident O2 Hanford-Livingston observing time (118 days) would imply a $p-$value of $\sim$20\% for this event, which is not statistically significant.

\vskip 4pt

In Fig.~\ref{fig:ifar} we show an alternative representation of our candidate triggers by plotting the cumulative distributions of candidate triggers for each observing run, and we compare them with an expected Poisson background. There we obtain the background triggers by redoing the Hanford-Livingston coincidence analysis on 2000 runs of unphysical timeslides, where each time slide is shifted by more than $\sim 10$ ms (the physical time delay between the two detectors). The top candidate, which is triggered in O2, is within $\sim 2 \sigma$ of the expected background. We therefore conclude that there is no statistically significant evidence for binary systems with large Love numbers in the data. In \S\ref{sec:rates_constraints} we translate our null detection to an upper limit on the merger rates of various models of exotic compact objects.

\vskip 4pt

While the IFARs of the candidate triggers in Table~\ref{tab:signalsFound_new} are not significant, it is nevertheless interesting that they are larger than the IFARs of our inspiral-only template triggers corresponding to previously detected BBH mergers; cf. Table~\ref{tab:signalsFound} below. This suggests that detailed modeling of the merger part of BSM compact objects would provide substantial improvements in the statistical significance of putative new triggers. In the future, it would be interesting to test whether the SNRs and IFARs of the triggers in Table~\ref{tab:signalsFound_new} would improve significantly when full inspiral-merger-ringdown waveforms are used for BSM compact object coalescences, though such efforts would necessarily be model-dependent and perhaps involve numerical simulations~\cite{Bezares:2018qwa, Bezares:2022obu, Croft:2022bxq, Bamber:2022pbs, Siemonsen:2023hko}. 

\vskip 4pt

Furthermore, in standard BBH searches the interpretation of marginal events is often supplemented by an estimated probability of whether the event is of astrophysical origin, $p_{\rm astro}$~\cite{pastro, Olsen:2022pin, Andres:2021vew}. This measure assumes a prior distribution for the intrinsic parameters and a source rate model, and have been used to improve the astrophysical interpretation of low IFAR marginal BBH events when $p_{\rm astro} > 0.5$. 
On the other hand, for binary systems with large $\tilde{\Lambda}$ the evaluated $p_{\rm astro}$ would likely be penalized if the commonly used prior distributions for BBH and BNS are used. Absent former detections of binaries with large $\tilde{\Lambda}$ it is also difficult to construct a well-informed source rate model for such systems.
It would be interesting to explore how $p_{\rm astro}$ can be adapted to make future searches more sensitive to a wider class of new GW signals.

\subsubsection{Known Binary Black Holes} \label{sec:known_bbhs}

One of the most interesting results of this work is the recovery of previously detected BBH merger signals within our template bank mass range in the O1--O3 Hanford and Livingston data. Crucially, while standard BBH searches rely on templates with full IMR waveforms, we show that one can recover the same signals with the inspiral-only TaylorF2 with Tides waveform in \S\ref{sec:waveform}, albeit with lower SNRs and IFARs. Additionally, the absence of spins in our templates contributes to SNR loss, especially for BBH events with large spins. Having said that, the IFAR estimates for zero-spin templates could improve for low-spin events, since the number of templates in the zero-spin template bank is smaller than that of non-zero spin template banks, and hence the look-elsewhere effect will be smaller even though the waveform match will be comparable. In what follows, we quantify the binary spin with the effective spin parameter $\chi_{\rm eff} = (m_1 \chi_1 + m_2 \chi_2) / M $.

\vskip 4pt

In Table~\ref{tab:signalsFound} we show the trigger details of known BBH signals. It is clear that all the best-fit templates have $\tilde{\Lambda} \lesssim \mathcal{O}(10^3 - 10^4)$, which is consistent with the fact that these events were detected by $\tilde{\Lambda} = 0 $ BBH template banks in earlier works, cf. Fig.~\ref{fig:effectualness}. In \S\ref{sec:bh_love} we show the parameter estimation results of these signals computed using the TaylorF2 with Tides model when including aligned-spin effects, and demonstrate that they are all indeed consistent with binaries that have zero Love numbers. Note that these events are triggered by either Bank 0, 1 or 2 because the parameter coverage of these three banks spans those relatively small values of $\tilde{\Lambda}$, as displayed in Fig.~\ref{fig:bank_params}. %

\vskip 4pt

To facilitate comparison between the inspiral-only search and standard IMR template searches, we duplicate the summary statistics of these signals reported in Refs.~\cite{Venumadhav:2019tad, Venumadhav:2019lyq, Olsen:2022pin, Ajit-O3b}, which use the same matched-filtering pipeline as this work. As expected, the SNRs for all events triggered by the inspiral-only template bank are smaller than those triggered by full IMR waveforms. Out of the 19 BBH triggers in Table~\ref{tab:signalsFound}:

\bgroup
\begin{table}[t!]
    \centering
    \setlength\tabcolsep{3pt}
    {\tabulinesep=1mm
    \begin{tabular}{|c||c|cc|cc|c||cc|c|}
        \hline \hline
         \multirow{3}{*}{Event Name} & \multicolumn{6}{c||}{TaylorF2 with Tides} & \multicolumn{3}{c|}{IMRPhenomD}  \\ & \multicolumn{6}{c||}{(Inspiral-only waveform, this work)} & \multicolumn{3}{c|}{ (Full IMR~\cite{Venumadhav:2019tad, Venumadhav:2019lyq, Olsen:2022pin, Ajit-O3b})}  \\\cline{2-10}
           & Bank & $\mathcal{M}$ [$M_\odot$] & $\tilde{\Lambda}$ &  $\rho_{\rm H}^2$ & $\rho_{\rm L}^2$ & IFAR$^{\ref{footnote:ifar}}$ [yr] & $\rho_{\rm H}^2$ & $\rho_{\rm L}^2$ & IFAR$^{\ref{footnote:ifar}}$ [yr] \\
         \hline
\rowcolor{gray!30} GW151012$^\dagger$ & \texttt{2} & 15.0 & 879 & 33.0 & 33.0 & 0.1 & 55.7 & 46.8 & $>1000$ \\ 
GW151226 & \texttt{1} & 8.4 & 1354 & 48.4 & 24.1 & 0.01 & 120.0 & 52.1 & $>1000$ \\
\rowcolor{gray!30} GW190412$^\dagger$ & \texttt{1} & 14.9 & 541 &  63.3 & 202.1 & $> 1000$ & 76.2 & 245.5 & $>1000$  \\
GW190503\_185404$^\dagger$ & \texttt{1} & 15.0 & 749 & 49.1 & 30.0 & 0.4 & 83.2 & 57.7 & $> 1000$  \\
\rowcolor{gray!30} GW190513\_205428$^\dagger$ & \texttt{1} & 15.0 & 749 & 40.9 & 40.3 & 44.7 & 78.0 & 66.0 & $>1000$  \\
GW190706\_222641$^\dagger$ & \texttt{2} & 15.0 & 879 & 58.5 & 53.8 & $> 1000$ & 91.3 & 79.2 & $> 1000$  \\
\rowcolor{gray!30} GW190707\_093326 & \texttt{0} & 9.8 & 868 & 55.2 & 77.8 & $> 1000$ & 63.7 & 97.5 & $> 1000$  \\
GW190720\_000836 & \texttt{1} & 10.1 & 2071 & 27.0 & 45.3 & 2.2 & 44.7 & 62.3 & $> 1000$  \\
\rowcolor{gray!30} GW190725\_174728 & \texttt{0} & 9.0 & 775 & 27.1 & 53.1 & 30.6 & 31.3 & 59.1 & 34.2   \\
GW190728\_064510 & \texttt{0} & 9.8 & 625 & 44.3 & 81.9 & $> 1000$  & 58.4 & 110.1 & $>1000$   \\
\rowcolor{gray!30} GW190828\_065509$^\dagger$ & \texttt{2} & 15.0 & 879  & 56.1 & 41.3 & $> 1000$  & 54.5 & 53.6 & $>1000$   \\
GW190915\_235702$^\dagger$ & \texttt{1} & 15.0 & 749 & 59.3 & 25.1 & 6.5  & 92.4 & 71.1 & $>1000$   \\
\rowcolor{gray!30} GW190930\_133541 & \texttt{1} & 9.5 & 2860 & 32.4 & 42.1 & 0.5  & 41.1 & 55.6 & $>1000$    \\ 
 GW191105\_143521 & \texttt{0} & 9.5 & 770 & 30.3 & 44.0 & 1.0  & 31.0 & 57.0 & $>1000$  \\
\rowcolor{gray!30} GW191129\_134029 & \texttt{0} & 8.41 & 530  & 55.6 & 87.0 & $> 1000$ & 73.1 & 95.1 & $>1000$  \\
GW191204\_171525 & \texttt{2} & 9.0 & 9655 & 32.1 & 124.6 & 5.6   &  87.8 & 183.3 & $>1000$  \\
\rowcolor{gray!30} GW191222\_033537$^\dagger$ & \texttt{1} &  15.0 & 749 & 65.3 & 29.4 & 75.4 & 73.7 & 66.7 & $>1000$   \\
 GW200225\_060421$^\dagger$ & \texttt{1} & 15.0 & 749  & 53.8 & 31.7 & 26.4  & 90.7 & 61.0 & $>1000$  \\
\rowcolor{gray!30} GW200316\_215755 & \texttt{1} & 10.4 & 1548  & 28.2 & 57.4 & 176 & 30.8 & 65.1 & 106  \\
         \hline \hline
    \end{tabular}}
    \caption{Coincident GW events in the O1$-$O3 Hanford$-$Livingston data that have been reported in earlier BBH searches and are recovered in this work. Recall that our template bank mass prior spans over $3 M_\odot < \mathcal{M} < 15 M_\odot$ and $M<40M_\odot$ (see Fig.~\ref{fig:bank_params}) --- we therefore only recover a subset of all known BBHs. Note that the best-fit templates of these events have $\tilde{\Lambda} \lesssim \mathcal{O}(10^3 - 10^4)$, which we showed in Fig.~\ref{fig:effectualness} is almost equivalent to the $\tilde{\Lambda} = 0$ BBH value.
    The events labeled by $^\dagger$ have source masses (as inferred from parameter estimation studies in earlier works) that are larger than our banks' upper $\mathcal{M}$ bound, and therefore have their best-fit templates cluster at the $\mathcal{M} = 15 M_\odot$ boundary. 
    For ease of comparison of our inspiral-only search with the results from full inspiral-merger-ringdown searches, we  show the corresponding statistics of the same signals reported in Refs.~\cite{Venumadhav:2019tad, Venumadhav:2019lyq, Olsen:2022pin, Ajit-O3b} (all of which use the same IAS search pipeline used in this work).}
    \label{tab:signalsFound}
\end{table}
\egroup

\begin{enumerate}

    \item Five events have no appreciable loss in IFARs despite a reduction in their measured SNRs of around $5-15$\%. These are GW190707\_093326, GW190725\_174728, GW190728\_064510,  GW191129\_134029 and GW200316\_215755. In most cases the reduced IFARs remain at the $> 1000$ per observing run level, implying that the coherent scores of these triggers still exceed the distribution of background scores generated by combining the Hanford and Livingston data at unphysical time slides~\cite{Venumadhav:2019tad}; 
    
    \item Nine of the events: GW151012, GW190412, GW190503\_185404, GW190513\_205428, GW190706\_222641, GW190828\_065509, GW190915\_235702, GW191222\_033537, and GW200225\_060421 are labeled by the $^\dagger$ superscript because their source masses, as inferred from comprehensive parameter estimation studies in earlier works~\cite{LIGOScientific:2018mvr, LIGOScientific:2020ibl, LIGOScientific:2021djp, Venumadhav:2019lyq, Venumadhav:2019tad, Olsen:2022pin, Nitz:2021zwj}, are larger than the upper bounds of the search mass region used in this work: $\mathcal{M} < 15 M_\odot$ and $M < 40 M_\odot$. Indeed, the best-fit templates of all these triggers lie at the $\mathcal{M} = 15 M_\odot$ mass boundary. It is often the case that a given signal is triggered by multiple banks and only the most significant trigger is reported by the pipeline; for these events the best trigger simply lies beyond our search space.
    It is therefore unsurprising that the $\rho^2$'s and IFARs of many of these events are appreciably lower than those reported in earlier works, where better-fitting IMR waveforms of BBH mergers at higher masses are used;

    \item The remaining five events have substantially decreased IFARs, depreciating what would have been identified as highly confident events to marginal ones. Of these, four also have substantially lower SNRs ($15-35$\% loss) than the search using the full IMRPhenomD model, which includes spins; these are GW151226, GW190720\_000836, GW190930\_133541, and GW191204\_171525. All of these systems show a preference for $\chi_{\rm eff} > 0$, excluding or nearly excluding $\chi_{\rm eff} = 0$ at the 90\% credible level when using IMR waveforms in parameter estimation~\cite{LIGOScientific:2021usb,LIGOScientific:2021djp}. This suggests that our search becomes increasingly insensitive as spins move away from $\chi_{\rm eff} \sim 0$, as expected. The remaining low IFAR event, GW191105\_143521, lost $\sim 15\%$ of its SNR in our search despite having low $\chi_{\rm eff}$. We attribute this loss in IFAR to changes in the distributions of background triggers compared to the full \texttt{IMRPhenomD} search.

\end{enumerate}

\noindent There are known BBH events whose masses fall within our search's mass range but which we did not recover due to a variety of well-understood reasons. As in previous works conducted using the same pipeline~\cite{Venumadhav:2019tad, Venumadhav:2019lyq, Olsen:2022pin, Ajit-O3b}, we only searched for coincident triggers in the Hanford and Livingston data but not in the Virgo data. As a result, our search did not recover GW190708\_232457, GW190814, and GW190925\_232845, which were detected via the Livingston-Virgo network~\cite{LIGOScientific:2021usb, Nitz:2021uxj}. We did not find GW170608~\cite{LIGOScientific:2018mvr}, as was previously the case in Ref.~\cite{Venumadhav:2019lyq}, because the Hanford data for this event was not provided in the earlier LVK bulk data release and therefore is not part of our coincidence search. The event GW190924\_021846~\cite{LIGOScientific:2021usb} was vetoed by the pipeline due to the presence of a loud glitch that occurred near the trigger~\cite{Olsen:2022pin}. Similarly, GW191216\_213338~\cite{LIGOScientific:2021djp} was vetoed because it exceeds the $\rho = 20$ threshold that is part of the pipeline's glitch mitigation procedure (see \cite{Venumadhav:2019tad, Ajit-O3b} for details). Finally, many of the marginal events that were reported in earlier works~\cite{Olsen:2022pin, Ajit-O3b, LIGOScientific:2021usb, LIGOScientific:2021djp, Nitz:2021zwj} are not detected in the present work. This is not surprising given that their statistical significance is already low when full IMR templates are used in the search. These marginal events include GW190704\_104834, GW190718\_160159, GW190821\_124821, GW190910\_012619, GW190917\_114630, GW190920\_113516, GW191103\_012549, GW191113\_071753, GW191126\_115259, GW191126\_115259, GW191219\_163120, GW191224\_043228, GW191228\_085854, GW200202\_154313, GW200210\_005122, GW200210\_092254, and GW200316\_235947.

\subsection{Parameter Estimation} \label{sec:pe}

To further examine the aforementioned detected BBH's, we perform Bayesian parameter estimation (PE) using our inspiral-only tidal waveform. Until this stage, we have been ignoring spins of BHs, but we now include aligned-spin contributions, via the orbit-aligned spin components $\chi_{1,2}$, to the phase evolution up to 3.5PN in (\ref{eqn:phase})~\cite{Arun:2004hn, Buonanno:2009zt}. This also helps us to more closely examine the degeneracy between spins and tidal effects. We use a 13-dimensional parameter space: six intrinsic parameters $\{\mathcal{M}, \eta, \chi_1, \chi_2, \Lambda_1,  \Lambda_2 \}$ and seven extrinsic parameters $\{D, t_c, \phi_c, \iota, \psi, \alpha, \delta \}$, where $D$ is the luminosity distance to the binary, $t_c$ and $\phi_c$ are respectively the time and phase of coalescence, $\iota$ is the inclination angle of the binary's orbital angular momentum with respect to the line-of-sight, $\psi$ is the polarization angle, $\alpha$ is right ascension, and $\delta$ is the declination.
Note that for the results presented here, we fix the dimensionless spin-induced quadrupole parameters $\kappa_i$ to those of black holes, $\kappa_i = 1$~\cite{Geroch:1970cd, Hansen:1974zz, Thorne:1980ru}.

\begin{table}[b!]
    \centering
    \begin{tabular}{|c|c|}
    \hline\hline

    Chirp mass, $\mathcal{M}$ & $\boldsymbol{U}[1, 25]\, \mathrm{M}_\odot$ \\ \hline
    Mass ratio, $q \equiv m_2 / m_1$ & $\boldsymbol{U}[0.05, 1]$ \\ \hline
    $\chi_1, \chi_2$ & $\boldsymbol{U}[-0.99, 0.99]$ \\ \hline
    $\Lambda_1, \Lambda_2$ & $\boldsymbol{U}[0, 10^4]$ \\ \hline
    $D$ (Uniform in Volume) & $[100, 3000]$\,Mpc \\ \hline
    $t_c$ & $\boldsymbol{U}[-0.2, 0.2]$ \\ \hline
    $\phi_c$ & $\boldsymbol{U}[0, 2\pi]$ \\ \hline
    Inclination Angle, $\cos(\iota)$ & $\boldsymbol{U}[-1, 1]$ \\ \hline
    Polarization Angle, $\psi$ & $\boldsymbol{U}[0, \pi]$ \\ \hline
    Right Ascension, $\alpha$ & $\boldsymbol{U}[0, 2\pi]$ \\ \hline
    Declination, $\sin(\delta)$ & $\boldsymbol{U}[-1, 1]$ \\ 
    \hline\hline
    \end{tabular}
    \caption{Priors for the 13 dimensional parameter space used for the parameter estimation results shown in Figs.~\ref{fig:corner}, \ref{fig:Love_constraint} and~\ref{fig:violin}.
    }
    \label{tab:priors}
\end{table}

\vskip 4pt

To remain consistent with our search, we only use the data from the Hanford and Livingston detectors. We use the standard Gaussian likelihood and assume that the noise is stationary at each detector over the relevant time scales and uncorrelated between detectors. 
We estimate the PSDs using Welch's method~\cite{1161901}. To efficiently and accurately sample the posteriors for each event, we utilize the code developed in Refs.~\cite{Wong:2023lgb, Edwards:2023sak} which uses a combination of relative binning~\cite{Zackay:2018qdy} (also known as the heterodyned likelihood estimation method~\cite{Cornish:2021lje}), hardware acceleration, differentiable waveforms, and normalizing flow enhanced sampling to perform minute time-scale PE. Finally, our priors are shown in Table~\ref{tab:priors}. For the intrinsic parameters, the priors are chosen to sufficiently cover the parameter space of our search with an additional buffer on either end (when possible) to ensure that the posteriors do not push against the boundaries. We discuss how these prior choices influence our results below. Additionally, we do not consider the events indicated with a dagger in Table~\ref{tab:signalsFound} since their true parameters lie outside the range considered in our search (see \S\ref{sec:known_bbhs}).

\vskip 4pt

The primary aims of our PE are to: \textit{i)} demonstrate that there is a mild degeneracy between the mass-weighted spin parameter $\chi_{\rm eff}$ and $\tilde{\Lambda}$ (\S\ref{sec:lambda-spin-degeneracy}); \textit{ii)} discuss the bias observed in the parameter inference from using an inspiral-only waveform, by comparing to parameter inferences carried out with IMR waveforms (\S\ref{sec:pe_bias}); \textit{iii)} measure the Love numbers of black holes (\S\ref{sec:bh_love}).

\subsubsection{$\mathcal{M}-\chi_{\rm eff}-\tilde{\Lambda}$ Degeneracy} \label{sec:lambda-spin-degeneracy}

As discussed in Section~\ref{sec:love_search}, we set the spin parameters to zero in the template bank waveforms for computational efficiency during the search. 
However, previous parameter inference performed on the BBH signals we detected typically allows for a range of spin values~\cite{LIGOScientific:2018mvr, LIGOScientific:2020ibl, LIGOScientific:2021usb, LIGOScientific:2021djp, Venumadhav:2019tad, Venumadhav:2019lyq, Olsen:2022pin, Ajit-O3b, Nitz:2018imz, Nitz:2019hdf, Nitz:2021uxj, Nitz:2021zwj}.
It is therefore interesting to investigate whether there exists a degeneracy between the spin parameters and $\tilde{\Lambda}$, and if so, whether it played a role in improving the effectiveness of our search.

\begin{figure}[b!]
    \centering    \includegraphics[width=0.68\textwidth]{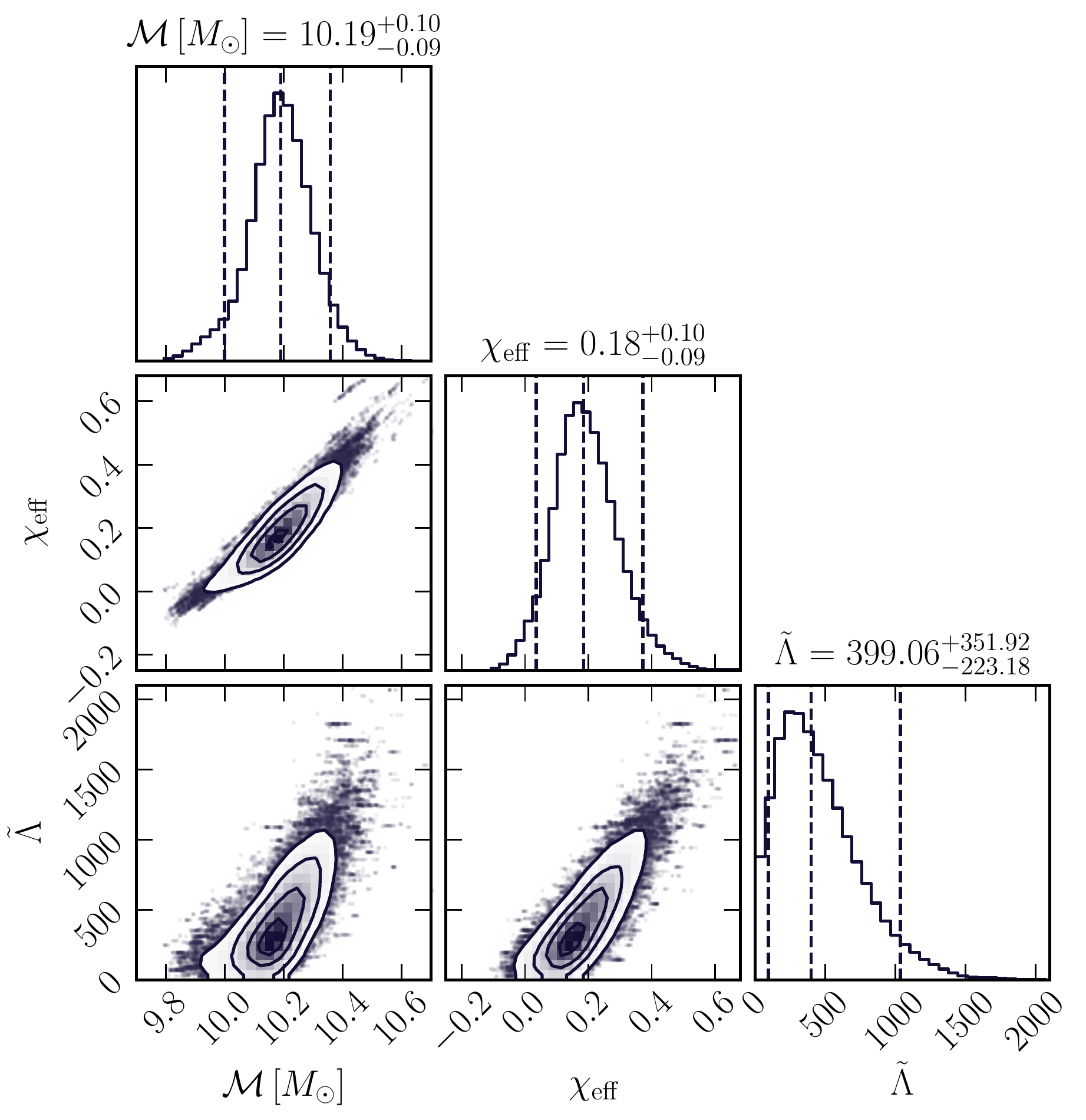}
    \caption{To highlight the degeneracy between $\tilde{\Lambda}$ and the other intrinsic parameters, we show the corner plot for one of the events: GW190707\_093326. Contours for the 2D panels correspond to 1$\sigma$, 2$\sigma$, and 3$\sigma$ regions respectively, whereas the titles for the marginalized posteriors quote the 90\% credible region. In particular, we can see that $\tilde{\Lambda} \geq 0$ leads to a larger estimation of the chirp mass. Similarly, $\tilde{\Lambda}$ and $\chi_{\rm eff}$ are positively correlated, meaning that our search region of $\tilde{\Lambda}\gtrsim 500$ can, to some extent, mimic waveforms with $\chi_{\rm eff}\lesssim 0$. Note that, due to our priors on $\Lambda_{1,2}$, we do not expect the $\tilde{\Lambda}$ posterior to peak at exactly zero, despite the sources being BBH (see \S\ref{sec:bh_love} for more discussion).}
    \label{fig:corner}
\end{figure}

\vskip 4pt

In Fig.~\ref{fig:corner}, we examine the degeneracy between $\chi_{\rm eff}$ and $\tilde{\Lambda}$ by looking at a corner plot of the posteriors in $\{\mathcal{M}, \chi_{\rm eff}, \tilde{\Lambda}\}$. Here we choose GW190707\_093326, as it clearly demonstrates the degeneracy in question, although similar results hold for all other sources. 
Fig.~\ref{fig:corner} illustrates that simultaneously increasing $\chi_{\rm eff}$ and $\tilde{\Lambda}$ leads to similar waveforms i.e., they are positively correlated. Furthermore, the degeneracy extends to $\mathcal{M}$ as well, with larger values of $\tilde{\Lambda}$ requiring larger values of chirp mass to maintain a similar waveform. Heuristically, one can understand the direction of this 3D degeneracy from the phase in PN theory~\cite{Arun:2004hn, Buonanno:2009zt, Flanagan:2007ix}: 
\beq
\begin{aligned}
    \psi_{\textrm{ins}}(f) & \supset  \frac{3}{128 \hskip 1pt ( \pi \mathcal{M} f)^{5/3}  } \left[ 1 + \cdots + \frac{113}{3} \hskip 1pt \chi  \hskip 1pt  v^3 + \cdots - \frac{39 \hskip 1pt \tilde{\Lambda}}{2} v^{10}\right] \, , \label{eqn:degeneracy_phase}
\end{aligned}
\eeq
where the reduced spin parameter $\chi$, which first appears at 1.5PN order, is usually dominated by  $\chi_{\rm eff}$~\cite{Kidder:1992fr, Poisson:1995ef, Ajith:2011ec}. 
Due to the negative sign on the tidal term in (\ref{eqn:degeneracy_phase}), lines of constant phase between the 1.5PN spin term and the 5PN tidal term appear positively correlated. 
The same argument applies to the degeneracy between $\tilde{\Lambda}$ and $\mathcal{M}$. 
The degeneracy between $\mathcal{M}$ and $\chi_{\rm eff}$ along the positive direction is well known: a binary which would have merged faster due to a larger chirp mass would be counterbalanced by the repulsive effect of having higher spins due to the ``orbital hang-up" phenomenon~\cite{Campanelli:2006uy, Roulet:2018jbe}.

\vskip 4pt

Interestingly, since our search space only covers $\tilde{\Lambda}\gtrsim 500$, our waveform can, to some extent, mimic waveforms in the negative $\chi_{\rm eff}$ region. We also verified that allowing for $\tilde{\Lambda}\leq 0$ results in posteriors extending further along the degeneracy in the $\chi_{\rm eff}\leq 0$ region. Nevertheless, since many of the detected BBH events in Table~\ref{tab:signalsFound} have the majority of their posterior support in the positive $\chi_{\rm eff}$ region~\cite{LIGOScientific:2021usb,LIGOScientific:2021djp}, the degeneracy discussed here is unlikely to have played a role in aiding the detection of those BBH events.

\vskip 4pt

One final thing to note here is that the details of the degeneracy discussed in this section may change if the object with large tidal deformability also has large spin-induced multipole moments. In particular, the spin-induced quadrupole, which first appears at 2PN~\cite{Poisson:1997ha}, can in some cases replace the tidal defomability as the dominant finite-size effect in the phase evolution of the binary (note though that extended objects generally have limited spins as they are bounded by their mass-shedding limit).
Future work should therefore look to include these effects in order to broaden the search space~\cite{Coogan:2022qxs, Chia:2022rwc} and explore the degeneracy in that larger dimensional space of intrinsic parameters.

\subsubsection{Tidal Waveform vs IMR Waveform} \label{sec:pe_bias}

Although our tidal waveform (\ref{eqn:total_wf}) is accurate for the inspiral region of the signal, the point-particle PN coefficients are not known up to arbitrary order. Full IMR waveforms such as \texttt{IMRPhenomXPHM}~\cite{Pratten:2020ceb} account for these unknown strong-gravity effects using semi-analytic models calibrated to numerical relativity merger simulations. In addition to the lack of a merger, the absence of other physical effects, such as precession and higher harmonics, makes our waveform model inaccurate to some degree. We therefore want to examine the bias introduced by performing PE on known BBH events using our tidal waveform. 

\vskip 4pt

\begin{figure}[t!]
    \centering
    \includegraphics[width=0.8\textwidth]{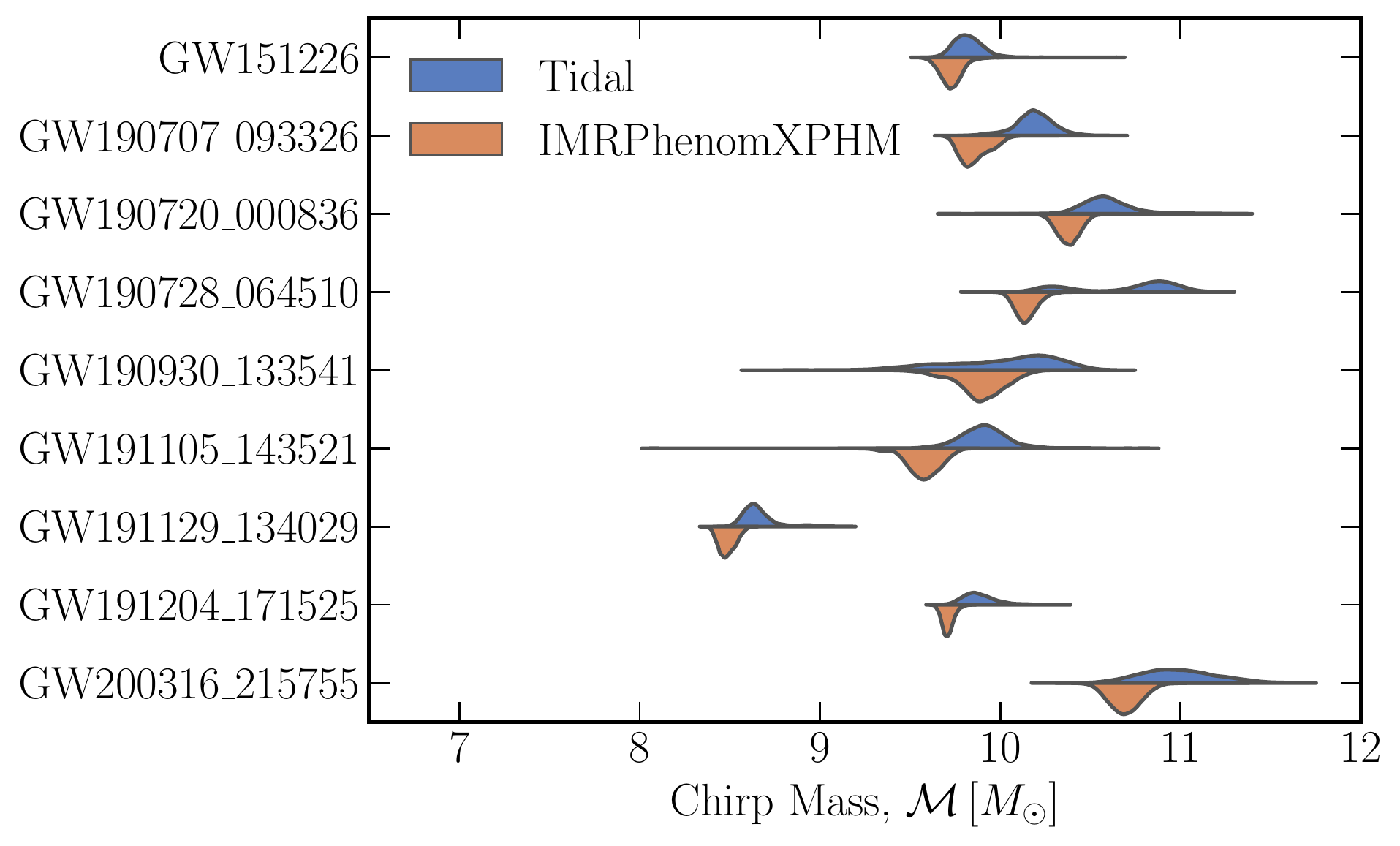}
    \caption{Violin plot showing the detector frame chirp mass $\mathcal{M}$ posteriors for all non-dagger events listed in Table~\ref{tab:signalsFound}. One can see that the inspiral-only tidal waveform systematically overestimates $\mathcal{M}$ compared to the state-of-the-art \texttt{IMRPhenomXPHM}. This can be partially explained by a combination of the missing higher-order phase contributions in our analytic tidal waveform; see (\ref{eqn:phase}), our $\Lambda_{1,2}$ prior choices, and the $\mathcal{M}-\tilde{\Lambda}$ degeneracy shown in Fig.~\ref{fig:corner}. }
    \label{fig:violin}
\end{figure}

In Fig.~\ref{fig:violin}, we show a violin plot for the non-dagger events from Table~\ref{tab:signalsFound}. In particular, we compare the marginalized posteriors for the detector frame chirp mass $\mathcal{M}$ obtained from the tidal waveform, which is evaluated with the PE code in Refs.~\cite{Wong:2023lgb, Edwards:2023sak}, and those obtained using the state-of-the-art \texttt{IMRPhenomXPHM} model~\cite{Pratten:2020ceb}, which were computed in earlier works~\cite{Venumadhav:2019tad, Venumadhav:2019lyq, Olsen:2022pin, Ajit-O3b} using \texttt{cogwheel}~\cite{Roulet:2022kot}. We focus on $\mathcal{M}$ because it is the dominant intrinsic parameter in (\ref{eqn:phase}) and should therefore be measured most precisely for systems where the inspiral contributes appreciably to the SNR. Consequently, its measurement is relatively insensitive to differences in the priors used in the two PE codes~\cite{Wong:2023lgb, Edwards:2023sak, Roulet:2022kot}. As can be seen from Fig.~\ref{fig:violin}, using the tidal waveform consistently overestimates $\mathcal{M}$. In particular, we find that the median chirp mass inferred by \texttt{IMRPhenomXPHM} is typically 1--6\% smaller than the value inferred by the tidal waveform.

\vskip 4pt

This bias can be partially understood from the degeneracy between $\mathcal{M} $ and $\tilde{\Lambda}$, illustrated in Fig.~\ref{fig:corner} for GW190707\_09332. In particular, we see that larger $\tilde{\Lambda}$ also leads to a higher $\mathcal{M}$ measurement. Indeed, we find that performing the same PE run on GW190707\_09332 but with $\Lambda_{1,2}\sim 0$ reduces the relative error on the median inferred chirp mass from $\sim3.5$\% to $\sim2$\%. The remaining differences stem from the inaccuracy of the inspiral-only tidal model at higher frequencies, where missing point-particle PN contributions are important. To check that this is indeed the origin of the bias, in addition to setting $\Lambda_{1,2}\sim0$, we only consider frequencies up to $160\,$Hz~\cite{Narikawa:2021pak}.  
Excluding frequencies above this bound from the computations ensures that the waveform is only evaluated where the inspiral PN expansion is known to be accurate. 
In this case, the relative error on $\mathcal{M}$ drops to $0.4$\%. 
We therefore conclude that using inspiral-only waveforms can lead to small but noticeable biases in parameter inference, and care must be taken when interpreting the inferred parameter values. 
Obtaining more precise inspiral waveforms with higher-order PN contributions would be highly beneficial for future analyses, as it would help mitigate this problem; see e.g. Refs.~\cite{Favata:2013rwa,Wade:2014vqa} for the influence of unknown PN terms on the measurement of $\tilde \Lambda$. 

\vskip 4pt

Finally, it is interesting to find that the posterior for GW190728\_064510 is multi-modal. Multi-modal posteriors for intrinsic parameters are not uncommon in the GW PE literature~\cite{Abbott:2018wiz, Nitz:2020mga, Estelles:2021jnz, Chia:2021mxq, Olsen:2021qin, LIGOScientific:2021djp} and could arise due to a host of reasons, including potential systematic errors in waveform models, model degeneracies~\cite{Nitz:2020mga, Estelles:2021jnz, Chia:2021mxq, Mehta:2021fgz}, prior effects~\cite{LIGOScientific:2021djp}, low SNR~\cite{LIGOScientific:2021djp,Huang:2018tqd}, and nonstationary noise~\cite{Ashton:2021tvz}, among others. In our case, we found that the lower-mass mode continues the trend (noted above) of a small systematic overestimate of $\mathcal{M}$ compared to \texttt{IMRPhenomXPHM}. On the other hand, the higher-mass peak is accompanied by a significantly higher $\chi_{\rm eff}$ value. Interestingly, this is the only event where the $\tilde{\Lambda}$ posterior is centered more than 3$\sigma$ from the $\tilde{\Lambda}=0$ boundary. It is therefore interesting that the interplay between $\mathcal{M}-\chi_{\rm eff}-\tilde{\Lambda}$ allow for this additional mode to be equally preferred by the data. Having said that, fixing $\tilde{\Lambda}=0$ almost entirely removes the posterior support for the high-mass mode and agrees similarly well with \texttt{IMRPhenomXPHM}.

\pagebreak

\section{Implications for Astrophysics} \label{sec:astro_implications}

In this section, we briefly discuss some implications of our results on known astrophysical GW sources, i.e. black holes and neutron stars.

\subsection{Love Number of Black Holes} \label{sec:bh_love}

The vanishing Love numbers of black holes~\cite{Chia:2020yla, Charalambous:2021mea, Kol:2011vg, Binnington:2009bb, Damour:2009vw} is a fundamental property of the Kerr solution in GR~\cite{Hui:2020xxx, Charalambous:2021kcz, Hui:2021vcv, Bonelli:2021uvf, Hui:2022vbh, Charalambous:2022rre, Ivanov:2022qqt}. This makes a measured deviation of $\tilde{\Lambda}$ from the BBH value not only a powerful probe for the existence of new types of compact objects but also an interesting test of GR in the strong-gravity regime. 

\vskip 4pt

\begin{figure}[b!]
    \centering
    \includegraphics[width=0.9\textwidth]{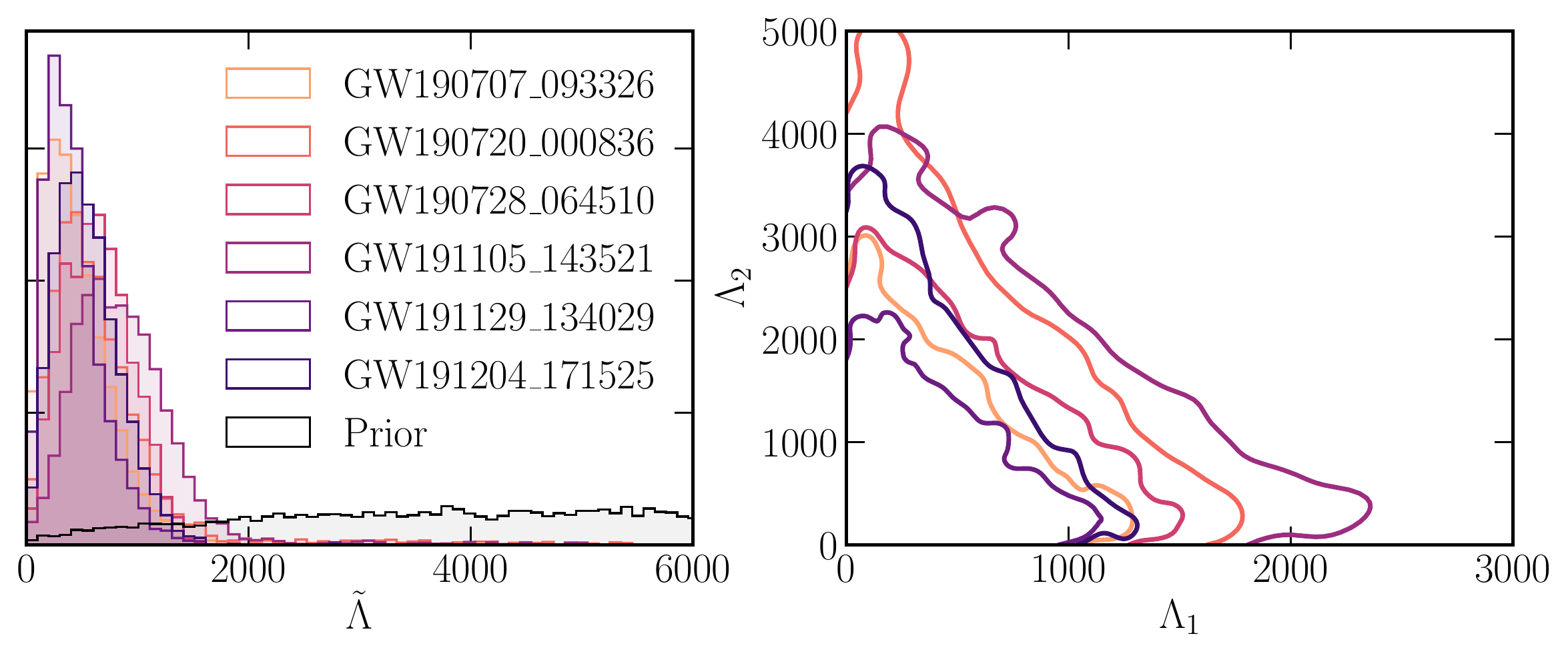}
    \caption{\textit{(left)} Marginalized $\tilde{\Lambda}$ posteriors for the six best non-dagger events listed in Table~\ref{tab:signalsFound}. We emphasize that the posteriors are prior dominated at $\tilde{\Lambda}\sim 0$, and therefore these results do not suggest evidence for $\tilde{\Lambda}> 0$ (see text for a discussion). \textit{(right)} 90\% credible region constraints on the individual Love numbers for the same six events. We note that since by definition $m_1 > m_2$, the constraint on $\Lambda_1$ is significantly tighter, cf. \eqref{eqn:lambdatilde}. }
    \label{fig:Love_constraint}
\end{figure}

On the left panel of Fig.~\ref{fig:Love_constraint} we show the $\tilde{\Lambda}$ marginalized posteriors for the six best non-dagger events from Table~\ref{tab:signalsFound} (all other events provide looser constraints). For these events we find that $\tilde{\Lambda} \lesssim 10^4$ at the 90\% credible interval, which is broadly consistent with the results found in Ref.~\cite{Narikawa:2021pak} which conducted a similar analysis. The Love number of black holes is therefore not well constrained by current GW detectors, though future detectors would improve the measurement precision by at least an order of magnitude~\cite{Cardoso:2017cfl, Puecher:2023twf}. The fact that $\tilde{\Lambda}$'s are constrained to be $\lesssim {\rm few} \times \mathcal{O}(10^3)$ can be traced back to Fig.~\ref{fig:effectualness} where we showed that BBH waveform ($\tilde{\Lambda}=0$) begin to significantly differ from the tidal waveform when $\tilde{\Lambda} \gtrsim \mathcal{O}(10^3)$.

\vskip 4pt

Note that the $\tilde{\Lambda}$ posteriors in Fig.~\ref{fig:Love_constraint} do not center around zero --- this is not a physical effect but arises entirely from the our choice of priors for the component mass and Love number in Table~\ref{tab:priors}. Specifically, our choice of $\Lambda_{1,2} \sim \boldsymbol{U}[0, 10^4]$, which ignores negative Love numbers as we consider such systems to be unnatural, results in a prior for $\tilde{\Lambda}$ that has vanishing support at $\tilde{\Lambda} = 0$. 
The posteriors in $\tilde{\Lambda}$ therefore consists of two regimes: 
\textit{i)} at $\tilde{\Lambda}\sim 0$ the posteriors fall to zero due to the lack of prior support as one approaches the boundary; \textit{ii)} for $\tilde{\Lambda}\gtrsim 10^3$ the data becomes informative (i.e., we are no longer prior dominated) and therefore the posteriors have no support in this region. Combined, these two effects lead to a peak in the $\tilde{\Lambda}$ posteriors away from zero.

\vskip 4pt
Since the dominant tidal term is proportional to the mass-weighted parameter $\tilde{\Lambda}$, cf. (\ref{eqn:phase}), the constraints on the individual $\Lambda_{1,2}$ are weaker. This can be seen on the right panel of Fig.~\ref{fig:Love_constraint} where we plot the 90\% credible region constraint for the same six events. Similar to $\tilde{\Lambda}$, these events all constrain the individual BH Love numbers to be $\lesssim {\rm few} \times \mathcal{O}(10^3)$. Note that $\Lambda_{1}$ is more constrained than $\Lambda_{2}$ because the heavier binary component provides a larger contribution to $\tilde{\Lambda}$ and is therefore more precisely measured.

\vskip 4pt

Finally, the differences between our posteriors for $\tilde{\Lambda}$ and those found in Ref.~\cite{Narikawa:2021pak} can be attributed predominantly to our different choices of $\Lambda_{1,2}$ priors. In particular, we were able to reproduce the results of Ref.~\cite{Narikawa:2021pak} when we repeated the analysis but with a prior that includes $\Lambda_{1,2} < 0 $. A prior which includes negative $\Lambda_{1,2}$ would certainly avoid the issue of vanishing prior support at $\tilde{\Lambda}=0$ described above, though astrophysical bodies with negative Love numbers would ``contract" instead of ``bulge" due to the gravitational pull exerted by the binary companion, which we consider to be unnatural. Astrophysical bodies can in principle attain negative Love numbers if they have negative pressure at their interiors, though such solutions are unlikely to be stable. To avoid spurious support for nonzero $\tilde \Lambda$, in the future it would be beneficial to use a prior that is uniform in $\tilde{\Lambda}$; see e.g. Refs.~\cite{Huang:2020ysn, Roulet:2021hcu} for a similar approach for spin measurements, whereby a uniform-in-$\chi_{\rm eff}$ prior is used instead of the uniform in component spin prior.

\subsection{Subsolar-mass Neutron Star Searches}

Neutron stars in merging binaries are the second most abundant source of GWs observed by the LIGO and Virgo detectors~\cite{TheLIGOScientific:2017qsa, LIGOScientific:2021usb}. It is important to emphasize that all searches for BNSs and neutron star-black hole binaries to date have been performed using BBH template banks. 
For instance, GW170817, the most confidently detected BNS thus far, was triggered by a BBH waveform template~\cite{TheLIGOScientific:2017qsa} --- a fact that already suggests that its effective tidal deformability is $\tilde{\Lambda} \lesssim 10^3$. Indeed, a comprehensive parameter estimation study inferred the event's component neutron stars to have masses $>1M_\odot$ and the Love numbers constrained to $\Lambda \lesssim 10^3$ at the $90\%$ credible level~\cite{TheLIGOScientific:2017qsa, Abbott:2018wiz}. From our effectualness study in Fig.~\ref{fig:effectualness}, it is unsurprising that this system with a relatively small $\tilde{\Lambda}$ was captured by a BBH bank without appreciable loss in detector sensitivity.

\vskip 4pt

The neutron stars observed to date are found to have masses $\gtrsim 1 M_\odot$ (see e.g. Ref.~\cite{Lattimer:2012nd} for a review). However, a recent claim on the detection of a neutron star~\cite{2022NatAs...6.1444D} with mass $\sim 0.77 M_\odot$ has raised the intriguing possibility for the formation of subsolar-mass neutron stars in astrophysical environments. If accurate, this event would challenge the standard paradigm of neutron star formation through gravitational collapse~\cite{Lattimer:2012nd}. For instance, a recent supernova simulation~\cite{Woosley:2020mze} for the formation of BNSs, which incorporates varying metallicity and mass-loss effects into their stellar progenitor models, suggests that neutron stars in close binaries have a minimum mass of approximately $ 1.2 M_\odot$. 
However, if the proto-neutron star is rapidly rotating, 
fragmentation of the star due to a dynamical instability could produce subsolar-mass neutron star remnants~\cite{Berezinsky:1988qca, Imshennik, Popov:2006ki}, though in these cases the supernova explosions are believed to impart large kicks therefore reducing the possibility of forming merging binaries. 
Despite theoretical challenges and tentative observational evidence~\cite{2022NatAs...6.1444D}, it would nevertheless be interesting to search for subsolar-mass BNSs in the future, which if detected would certainly challenge the prevailing astrophysical framework.

\vskip 4pt

In order to search for subsolar-mass BNSs, one would have to incorporate the Love number in the template waveform. 
This is because subsolar-mass neutron stars are known to have Love numbers that are orders of magnitude larger than those of solar-mass neutron stars, reaching $\Lambda \sim 10^4 - 10^5$ when $m \sim 0.8 M_\odot$~\cite{Yagi:2013awa, Silva:2016myw}. 
Physically, this arises because the self-gravity of a lower mass neutron star is weaker, resulting in a stellar equilibrium configuration with a larger radius for a given neutron degeneracy pressure and matter pressure in the high-density core supporting the star. 
Subsolar-mass neutron stars therefore naturally attain such large values of Love numbers by virtue of the $\Lambda \propto (r/m)^5$ scaling. Interestingly, this behavior is relatively insensitive to the precise nuclear equation of state~\cite{Yagi:2013awa, Silva:2016myw} and is a general phenomenon for subsolar-mass neutron stars. Fig.~\ref{fig:effectualness} suggests that current searches, which only include BBH template banks, would potentially be insensitive to such signals with large values of $\tilde{\Lambda}$ (note however that we did not extend our effectualness study to the subsolar-mass regime in Fig.~\ref{fig:effectualness}). Indeed, a recent comprehensive investigation~\cite{Bandopadhyay:2022tbi} suggests that the loss in sensitive volume for subsolar-mass neutron stars could degrade as much as $\sim 70 \%$ when subsolar-mass BBH template banks are used in the search~\cite{LIGOScientific:2018glc, LIGOScientific:2021job, Nitz:2021vqh, Nitz:2022ltl}.

\vskip 4pt

In this work, we restricted our search to the $\mathcal{M} > 3 M_\odot$ BBH mass range because the number of templates, and therefore the computational cost in the search, increases rapidly with decreasing chirp mass. 
Intuitively, this is the case because the lower the chirp mass, the larger the number of orbiting cycles the LIGO and Virgo detectors would observe, hence the more sensitive the detectors are to the phase coherence between the templates and the GW signals. We estimated that reducing the minimum chirp mass of our search space from $3 M_\odot$ to $0.1 M_\odot$ would increase the number of templates in our bank from $2.8 \times 10^4$ to more than $10^6$, which would be computationally challenging. As a result of the $\mathcal{M} > 3 M_\odot$ lower bound and our minimum cutoff requirement of $f_c > 60$ Hz, the range in Love numbers that our search covers is bounded by $\tilde{\Lambda} \lesssim 4 \times 10^6$; see Fig.~\ref{fig:bank_params}. 
An extension of our search space towards the low-mass region would naturally accommodate the $\tilde{\Lambda} \gg 10^6$ regime, and is well suited to search for the presence of subsolar-mass neutron stars. We hope to pursue this interesting line of research in future work.

\pagebreak

\section{Implications for New Physics} \label{sec:bsm_implications}

In this section we translate our null detection to model-independent constraints on exotic compact object mergers (\S\ref{sec:bsm_implications_model_independent}). In addition, we discuss how these constraints are mapped to bounds on a few specific models of BSM compact objects (\S\ref{sec:bsm_implications_dmmodels}).

\subsection{Model Independent Constraints} \label{sec:bsm_implications_model_independent}

The null detection of binary systems with large Love numbers places an upper limit on the rates at which these binaries would merge in the LIGO-Virgo observation bands. We provide such an estimate in \S\ref{sec:rates_constraints}. In \S\ref{sec:love_constraints}, we map our search parameter space in Fig.~\ref{fig:bank_params} into an approximate region in the BSM compact objects' parameter space for which these rate constraints apply.

\subsubsection{Upper Limit on Merger Rates} \label{sec:rates_constraints}

We can use our null detection of binaries with large Love numbers to place limits on the rate of such binary mergers. We assign an upper limit at 90\% confidence using the loudest event method~\cite{Bis09}, which assumes the event is modeled as a rare Poisson process. With the loudest trigger interpreted as noise, the rate limit is given as~\cite{Bis09}
\begin{equation}
R_{90, i} = \frac{2.3}{ \langle V T \rangle_i } \, ,
\end{equation}
where $V$ is the estimated sensitive volume of the analysis to a chosen source population assessed at the false alarm rate of the most significant observed candidate, $T$ is the length of the Hanford-Livingston coincidence observation period and the angular bracket denotes averaging over the $i-$th bin in intrinsic parameter space.

\vskip 4pt

In principle, a robust estimation of $V$ involves injecting a simulated source population into real data and evaluating the detection pipeline's efficiency as a function of the distance to a source and its intrinsic parameters; see e.g. Refs.~\cite{LIGOScientific:2018glc, Nitz:2019hdf} for such computations performed for subsolar-mass black hole binary rate constraints. In this work, we adopt a simplified approach whereby we use the root-mean-square SNR of the loudest trigger in Table~\ref{tab:signalsFound_new} (GPS time of 1170911358.28) as the threshold for the detector sensitivity, $\rho^2_{\rm thres}= (\rho^2_{\rm Hanford} + \rho^2_{\rm Livingston}) / 2 = 38.6$.
From the definition of the SNR and using the waveform model in \S\ref{sec:waveform_model}, 
\begin{equation}
    \rho_i \propto \frac{\mathcal{M}_{i}^{5/6}}{D_i} \left[ \int \d f  \, \frac{f^{-7/3}}{S_n(f)} \, \theta (f_{c, i} - f)  \right]^{1/2} \, , \label{eqn:snr_i}
\end{equation}
the minimum threshold in $\rho$ sets the maximum distance at which the source is detectable. We then simulate the SNRs of the signal population by fixing the source distance at $1$ Gpc but randomly sample over the distributions of masses and Love numbers that were used to build the template bank in Fig.~\ref{fig:bank_params}. 
The volumetric fraction of signals that would be retained is then estimated to be $(\rho_{i, 1\;\text{Gpc}} / \rho_{\rm thres} )^3$, which arises due to our step function approximation for detectable $\rho$, with the threshold located at $\rho_{\rm thres}$. The resulting $VT$ estimate is given as
\begin{equation}
 \frac{\langle V T \rangle_i}{\mathrm{Gpc}^3 \, \mathrm{yr}} \approx \frac{4 \pi }{3} \left\langle \left(\frac{ \rho_{i, 1\;\text{Gpc}}}{ \rho_{\rm thres}} \right)^3  \right\rangle \left(  \frac{1}{2.26} \right)^3  \left( \frac{T}{\mathrm{yr}} \right) \, ,  \label{eqn:v_estimate}
\end{equation}
where a correction factor of $(128/25)^{1/2} = 2.26$ to the horizon distance is included to take into account the effect of averaging over the detectors' angular response, the binary orbital orientation, and GW polarization angle~\cite{Finn:1992xs}. Since the three LIGO-Virgo observing runs have different characteristic noise curves and coincident observing periods, we compute $\langle VT \rangle_i$ separately for each run and sum them to obtain the total volume-time.

\vskip 4pt

\begin{figure}[t!]
    \centering
    \includegraphics[width=0.6\textwidth]{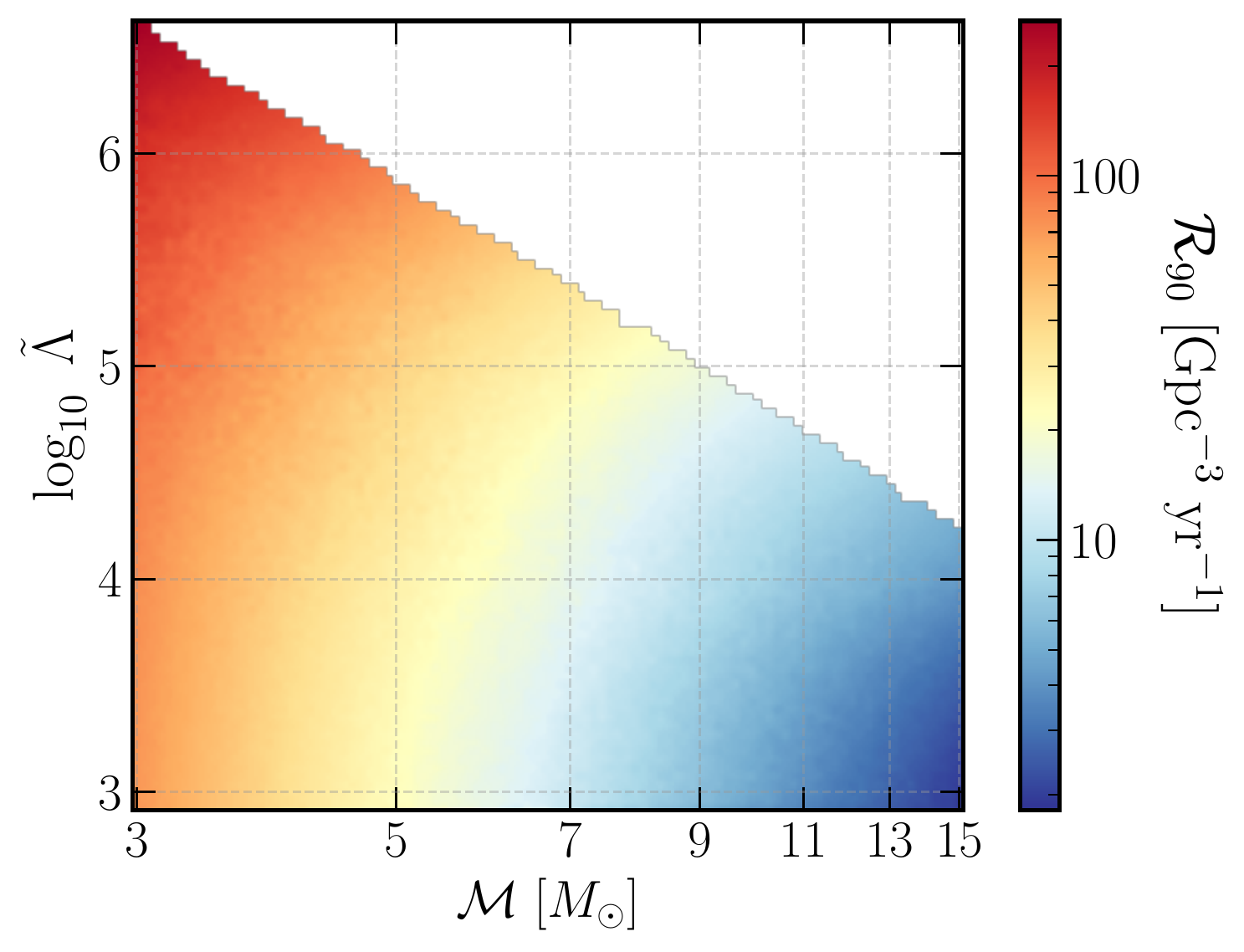}
    \caption{We use our null detection to put constraints on the merger rates of binaries with large $\tilde{\Lambda}$. Using the IFAR of our most significant event, we derive a 90\% CL upper limit on the merger rate as a function of detector-frame chirp mass and the leading-order tidal parameter of the binary (cf. the loudest event method \cite{Bis09}). The upper-right region is not colored as it is beyond the current extent of our template banks. }
    \label{fig:vt}
\end{figure}

Figure~\ref{fig:vt} shows the approximate $90\%$ confidence upper limit for the exotic binary merger rate, which fall in the $\sim 1 - 300$ Gpc$^{-3}$ yr$^{-1}$ range depending on binary parameters. For comparison, the merger rates of known BBHs in the same chirp mass range are approximately $\mathcal{R} \sim 1 - 50$ Gpc$^{-3}$ yr$^{-1}$~\cite{KAGRA:2021duu}. The constraints become stronger as we move from small to large values of $\mathcal{M}$ because the putative signal would have been louder. In particular, due to the $\rho \propto \mathcal{M}^{5/6}$ scaling in (\ref{eqn:snr_i}) the constraint scales as $R_{90} \propto \mathcal{M}^{-15/6}$. On the other hand, the constraints become weaker as we move from smaller to larger values of $\tilde{\Lambda}$ as the cutoff frequency of the waveform model decreases with increasing $\tilde{\Lambda}$. Systems with larger values of $\tilde{\Lambda}$ therefore lead to smaller amounts of SNR integrated in the data and correspondingly a smaller sensitive volume. Although the analytic scaling between $R_{90}$ and $\tilde{\Lambda}$ is less straight-forward to derive, we deduce empirically $R_{90} \propto \tilde{\Lambda}^{1/10}$ to be a good approximation. The upper limit rate is therefore much more sensitive to scalings with $\mathcal{M}$ than with $\tilde{\Lambda}$. Finally it is interesting that, despite the crude estimate (\ref{eqn:v_estimate}), our rate constraints of $R_{90} \sim 10^2$ Gpc${}^{-3}$ yr${}^{-1}$ at $\mathcal{M}=3 M_\odot$ and $\tilde{\Lambda} \to 0$ is broadly consistent with the rate constraints obtained for subsolar-mass BBHs which were computed with full injection campaigns involving the real data~\cite{LIGOScientific:2021job, Nitz:2022ltl}.

\subsubsection{Love Number Parameter Space} \label{sec:love_constraints}

The search parameter space in Fig.~\ref{fig:bank_params} and the merger rate constraints in Fig.~\ref{fig:vt} are shown as a function of $\tilde{\Lambda}$, $M$ and $\mathcal{M}$, which are parameters of the binary system. In order to express the constraints as a function of the intrinsic parameters of individual sources, i.e. in terms of $m_1$, $m_2$, $\Lambda_1$ and $\Lambda_2$, one would have to make an assumption on the nature of the binary system. For instance, one simplified assumption commonly used in the literature is to assume that both of the binary components have equal masses and are the same type of compact object. 

\vskip 4pt

In this work we assume that the binary is composed of a  black hole and a large-$\Lambda$ object. We believe this scenario is more astrophysically likely, partly because black holes are already known to exist and are abundant in Nature. In addition, our search mass coverage of $3M_\odot < \mathcal{M} < 15 M_\odot$ and $M < 40 M_\odot$ spans over black hole masses -- other astrophysical sources such as neutron stars have maximum masses that are below $3 M_\odot$, making them unlikely to have significant overlap with our search space. Furthermore, this assumption allows us to set either the Love number of the heavier component $\Lambda_1$ or that of the lighter component $\Lambda_2$ to zero,  thereby providing a one-to-one mapping between the effective binary Love number $\tilde{\Lambda}$ in (\ref{eqn:lambdatilde}) and the Love number of the binary component.

\begin{figure}[t!]
    \centering
    \includegraphics[width=\textwidth, trim=0 0 0 0]{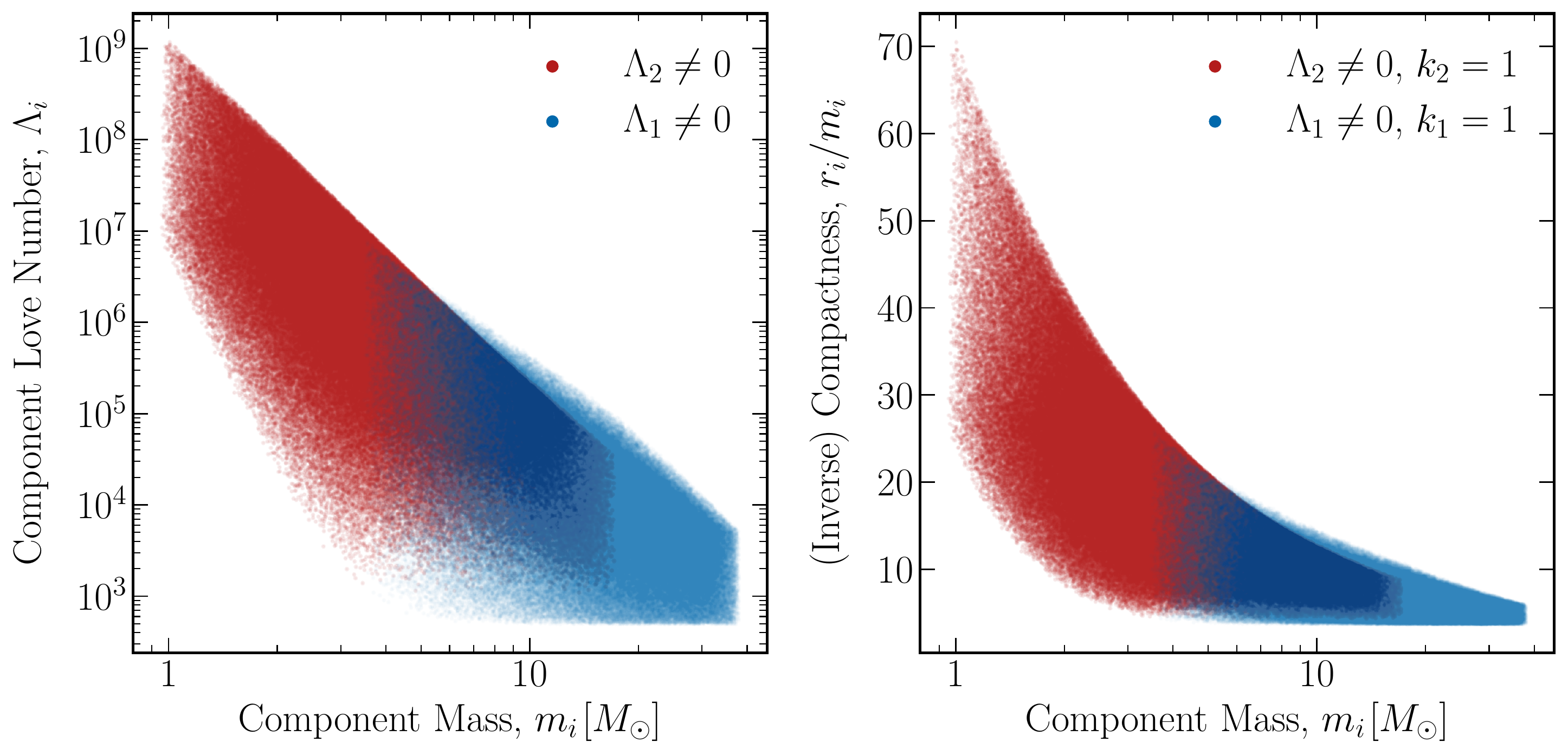}
    \caption{We map the binary parameter space $(M, \tilde{\Lambda})$ probed in our search (shown in Fig.~\ref{fig:bank_params}) to the parameter space of the individual objects. We assume that the binary consists of a black hole ($\Lambda=0$) and an exotic object with a non-vanishing Love number, as such a scenario is more likely than both binary components being exotic. \textit{(left)} The parameter space of the exotic object in the binary to which our merger rate constraints apply. The red region denotes the constraints applied when the lighter binary component (with mass $m_2$) is the exotic object while the heavier component (with mass $m_1$) is a black hole; the converse applies to the blue region. \textit{(right)} Same as the left panel, except we show the constraints in terms of the object's (inverse) compactness. Here we use the relationship $\Lambda \propto k (r/m)^5$ in (\ref{eqn:tides}) and set the constant $k = 1$ as a fiducial guide. The left panel is in general the more instructive plot since the radius and $k$ of a general BSM compact object, such as those with smoothly-varying density distributions, are not well defined (see also Footnote~\ref{ref:k2_illdefined}).} 
    \label{fig:param_bound}
\end{figure}

\vskip 4pt

In Fig.~\ref{fig:param_bound} we show the ranges of parameter space of a BSM compact object
for which the constraints depicted in Fig.~\ref{fig:vt} apply. On the left panel, we show the constraints in terms of the Love number of the object, either when it is the lighter component (red) or when it is the heavier component (blue). The red region encompasses much larger values of component Love numbers and therefore offers wider constraints. The red region is also the more astrophysically-realistic scenario since black holes, which do not have a maximum mass, are more likely to have formed the heavier binary component. In \S\ref{sec:bsm_implications_dmmodels} we will use the red region as our default parameter region for constraining several models of BSM objects. An exception is made for the gravitational atom, whereby a bosonic configuration grows spontaneously around a rotating black hole -- in this scenario we shall use the blue region for more conservative constraints. 

\vskip 4pt

In the right panel of Fig.~\ref{fig:param_bound} we use the relation between $\Lambda$ and $r/m$ in (\ref{eqn:tides}) to translate the range on the left panel to the (inverse) compactness of the exotic compact object. As a fiducial guide, we take $k=1$ for the dimensionless second Love number -- for other values of $k$ one could adjust the vertical axis according to the $(r/m) \propto k^{-1/5}$ scaling. The panel shows that for low-mass systems, our null detection constrains objects with sizes approximately in the range $5 \lesssim r/m \lesssim 70$. It should be emphasized, however, that many proposed objects in BSM scenarios do not have a well-defined radius. This typically occurs when the compact object has a smoothly-varying density distribution which formally vanishes at spatial infinity. In these cases, the left panel of Fig.~\ref{fig:param_bound} is the more instructive constraints plot.

\subsection{Bounds on Beyond Standard Model Physics} \label{sec:bsm_implications_dmmodels}

The parameter space displayed in Fig.~\ref{fig:param_bound}, for which our rates constraints apply, does not assume a specific model of exotic compact object. In this section we discuss how Fig.~\ref{fig:param_bound} translates to the model parameter space of a few representative examples of exotic compact objects proposed in the literature. Our discussion is by no means comprehensive and serves only as guide on the approximate region of model space that is constrained by our null detection, see e.g. Refs.~\cite{Giudice:2016zpa, Deliyergiyev:2019vti} for more examples of exotic compact objects discussed in the literature. 

\vskip 4pt

Motivated by the indisputable evidence for the existence of dark matter, virtually all BSM proposals posit the existence of at least a new degree of freedom in our Universe. In this section we separate our discussion into \S\ref{sec:ultralight_dm} and \S\ref{sec:particle_dm} -- the former describing exotic compact objects formed from ultralight boson fields~\cite{Hu:2000ke, Hui:2016ltb, Hui:2021tkt} and the latter formed from particle dark matter~\cite{Bertone:2004pz, Arcadi:2017kky}. The two classes of BSM physics are differentiated depending on if they display coherent wave-like properties at astrophysical scales. As a rough fiducial guide, we consider ultralight bosons to have masses approximately $\mu \lesssim 10^{-10}$ eV, as their Compton wavelengths of $\gtrsim 1$ km are comparable or larger than the sizes of black holes. Conversely, we consider particle dark matter to have masses $\mu \gtrsim 10^{-10}$ eV, though the models we focus on have $\mu \gtrsim \mathcal{O}$(MeV).

\subsubsection{Ultralight Bosons} \label{sec:ultralight_dm}

We consider the \textit{mini boson star} and the \textit{gravitational atom} as representative examples of compact objects formed from ultralight boson field. We shall see that our constraints apply to ultralight boson masses of the order of $\sim 10^{-12} - 10^{-11}$ eV for these types of compact objects. 

\vskip 4pt

\subsubsection*{Mini Boson Stars} 

Boson stars~\cite{Kaup1968, Ruffini1969, Breit:1983nr, Jetzer:1991jr, Schunck:2003kk, Colpi1986, Liebling:2012fv} constitute a generic class of BSM objects that arise from a new $U(1)$ complex scalar field. They are prototypical examples of BSM compact objects as they are relatively simple systems with many of their properties well explored in the literature. A key feature of boson stars is that they have regular boundary conditions at their centers --- in fact, many of the properties of boson stars are determined by the scalar field value at the origin, $\Phi_0$. The larger the central values, the more important relativistic and strong gravity effects are, and the more compact the boson stars. 

\vskip 4pt

Mini boson stars~\cite{Kaup1968, Ruffini1969, Breit:1983nr, Jetzer:1991jr, Schunck:2003kk} form a subclass of boson stars in that they are only bound through self-gravity and do not contain self interaction operators in the scalar potential. The minimally coupled complex scalar field $\Phi$ is described by the action
\begin{equation}
S = \int d^4 x \sqrt{-g} \left[ \frac{R}{16\pi} - \nabla_\alpha \Phi^* \nabla^\alpha \Phi - \mu^2 |\Phi|^2 \right] \, ,   \label{eqn:mini_BS_action}
\end{equation}
where $R$ is the Ricci scalar and $\mu$ is the boson mass. 
The self-gravity of mini boson stars is balanced by the ``quantum pressure" which arise due to Heisenberg's uncertainty principle, resulting in ultralight particles which cannot be localized within distances shorter than their de Broglie wavelengths. 
Mini boson stars are therefore effectively macroscopic coherent wave objects which behave like a Bose-Einstein condensate of astrophysical scales.

\vskip 4pt

A boson field with a given mass $\mu$ admits a family of stable mini boson star configurations, with each solution characterized by their central values $\Phi_0$. As $\Phi_0$ increases, the boson star mass, $M_{\rm BS}$, increases as well. This trend proceeds until the central field value is sufficiently large to yield solutions that are unstable to linear perturbations. Like white dwarfs and neutron stars, mini boson stars also have a theoretical maximum mass which is given as~\cite{Breit:1983nr}
\begin{equation}
    M_{\rm BS, max} \approx 0.633 \frac{M_{\rm pl}^2}{\mu} = 8.4 M_\odot  \left( \frac{10^{-11} \text{ eV}}{\mu} \right) \, . \label{eqn:mini_bosonstar_Mmax}
\end{equation}
These boson stars are called ``mini" because their maximum mass scales more slowly with $\mu$ than the Chandrasekhar limit for white dwarfs due to electron degeneracy pressure, $M_{\rm WD, max} \sim M_{\rm pl}^3 / \mu^2_{\rm electron}$, and therefore generally have smaller masses.

\vskip 4pt

The Love numbers for mini boson stars, $\Lambda_{\rm BS}$, has been computed both analytically in the non-relativistic Newtonian limit and numerically in the relativistic regime~\cite{Mendes:2016vdr, Sennett:2017etc, Cardoso:2017cfl}. The result can be expressed succinctly as
\begin{equation}
    \Lambda_{\rm BS} \simeq 10^3 \left( \frac{ k_{\rm BS} (\Phi_0)}{10} \right) \left( \frac{0.633}{\alpha_{\rm BS}} \right)^{10} \, , \quad \alpha_{\rm BS} \equiv \frac{M_{\rm BS} \mu}{M_{\rm pl}^2}  \approx 0.633 \left( \frac{M_{\rm BS}}{M_{\rm BS, max}} \right) \, , \label{eqn:mini_BS_Lambda}
\end{equation}
where $k_{\rm BS}$ is the constant $k$ defined in (\ref{eqn:tides}) for boson stars and $\alpha_{\rm BS}$ is the ratio of the boson star mass to the Compton wavelength of the boson field. We substituted (\ref{eqn:mini_bosonstar_Mmax}) into the definition of $\alpha_{\rm BS}$ in (\ref{eqn:mini_BS_Lambda}) to illustrate that stable and hence physically motivated solutions have $\alpha_{\rm BS} < 0.633$. 
Crucially, we observe that $\Lambda_{\rm BS}$ scales very sensitively with $\alpha_{\rm BS}$ and therefore easily spans over several orders of magnitude. 
The dimensionless constant $k_{\rm BS}$ depends on the central value of the scalar field: as  $\Phi_0 \to 0$ the system approaches the non-relativistic Newtonian limit, with $k_{\rm BS} \approx 1149$~\cite{Mendes:2016vdr}; as $\Phi_0$ increases the system approaches the relativistic regime and the coefficient can be as small as $k_{\rm BS} \sim 10$~\cite{Mendes:2016vdr, Sennett:2017etc}.

\begin{figure}[t!]
    \centering
    \includegraphics[width=\textwidth, trim=0 0 0 0]{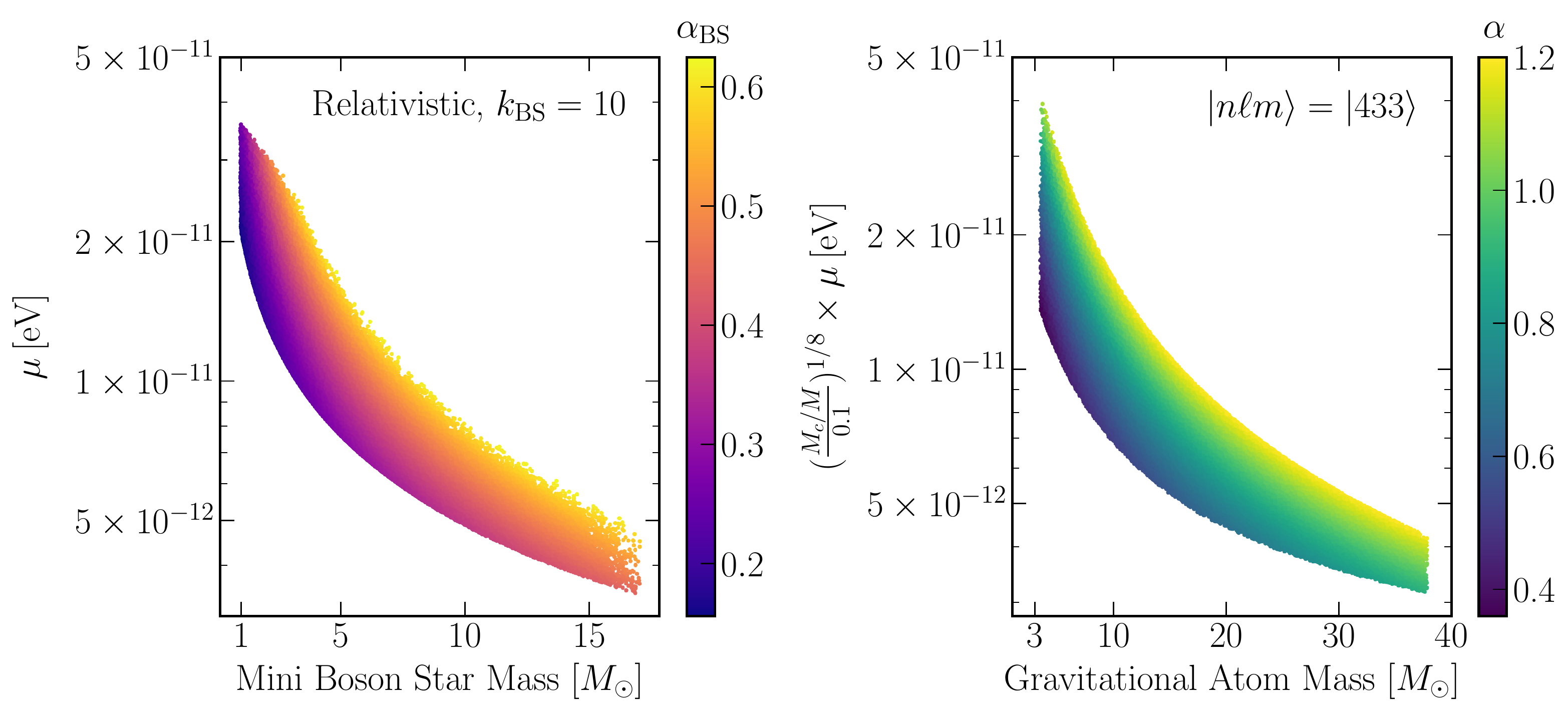}
    \caption{Same as Fig.~\ref{fig:param_bound}, except expressed in terms of 
    BSM physics parameter space corresponding to ultra-light bosons. \textit{(left)} Parameters of mini boson stars (BS), which couple minimally through gravity, to which our rate constraints apply. From the relationship $\Lambda_\mathrm{BS} \propto k_\mathrm{BS} (r/m)^5$, we use the tidal constant $k_{\rm BS}=10$ as the representative value for a relativistic BS. In the Newtonian non-relativistic limit, $k_{\rm BS} \to 1149$ and the vertical axis would scale mildly by a factor of $k_{\rm BS}^{1/10}$. The physically feasible regime corresponds to a bound on the ratio of the boson star mass to the Compton wavelength of the boson field: $\alpha_{\rm BS} \lesssim 0.633$. \textit{(right)} The parameter space constrained for a minimally-coupled scalar gravitational atom ($M_c$, $M$ are the masses of the boson cloud and black hole respectively). Here we show the constraints applied to the $\ket{n \ell m} = \ket{433}$ eigenstate; for other $\ell-$eigenstates, the vertical axis would change slightly with the tidal constant as $k^{-1/8}_{c, \ell} \sim \mathcal{O}(1)$, and the range of the gravitational fine structure constant $\alpha$ probed would be adjusted according to the $\alpha / \ell \lesssim 0.4$ physical region (see section~\ref{sec:ultralight_dm} for more discussion).} 
    \label{fig:ultralight_dm}
\end{figure}

\vskip 4pt

On the left panel of Fig.~\ref{fig:ultralight_dm} we show the parameter space of merging mini boson stars to which our rate constraints apply. Since mini boson star masses can span over both the black hole mass range and the subsolar-mass range, we assume they are the lighter component in the binary system (red region of Fig.~\ref{fig:param_bound}). In Fig.~\ref{fig:ultralight_dm} we use $k_{\rm BS}=10$ as the representative value for the relativistic case, though the dependence on $M_{\rm BS} \mu \propto k^{1/10}_{\rm BS}$ is rather weak. We see that, depending on the boson star mass, our constraints apply to ultralight bosons with masses over the range $10^{-12} - 10^{-11}$ eV. This range is primarily determined by the compactness of the star, which scales inversely with $\alpha_{\rm BS}$, and the finite frequency ranges at which ground-based detectors are capable of probing; cf. (\ref{eqn:lambda_bound}).

\vskip 4pt

\subsubsection*{Superradiant Clouds/ Gravitational Atoms} 

Another candidate for compact objects that arise in many BSM scenarios is the gravitational atom~\cite{Arvanitaki:2009fg, Arvanitaki:2010sy, Baumann:2019eav, Detweiler:1980uk, Dolan:2007mj}. The gravitational atom is a bosonic cloud configuration which would grow spontaneously around a rotating black hole through a process called black hole superradiance~\cite{Zeldovich:1971a, Starobinsky:1973aij, Bekenstein:1998nt, Brito:2015oca}. This process is triggered if \textit{i)} the Compton wavelength of the boson is comparable to the size of the black hole, and \textit{ii)} the initial black hole's spin is sufficiently high to satisfy the superradiance inequality~\cite{Zeldovich:1971a, Starobinsky:1973aij, Bekenstein:1998nt, Brito:2015oca}. These bound states are called gravitational atoms because they resemble the proton-electron structure of the hydrogen atom. Crucially, unlike boson stars, they do not have regular boundary conditions at the centre but are instead characterized by the purely-ingoing boundary condition at the black hole horizon. 

\vskip 4pt

Although black hole superradiance would be triggered by both minimally-coupled~\cite{Arvanitaki:2009fg, Arvanitaki:2010sy, Baumann:2019eav, Detweiler:1980uk, Dolan:2007mj} and self-interacting~\cite{Yoshino:2012kn, Baryakhtar:2020gao, Omiya:2022gwu, Chia:2022udn} boson fields, we shall focus on the minimally-coupled case as this is the simplest and most explored scenario in the literature. For simplicity, we shall also limit ourselves to ultralight scalar fields, though we do not expect our quantitative results to differ substantially for ultralight vector and tensor fields. The action of the gravitational atom which we investigate is therefore the same as (\ref{eqn:mini_BS_action}) for a complex scalar field and can be trivially extended to the real scalar field. Crucially, the background spacetime $g_{\alpha \beta}$ in the Fermi frame of the gravitational atom would now be the Kerr metric.

\vskip 4pt

The eigenstates of the scalar gravitational atom are determined by the set of ``quantum numbers" $\{n, \ell, m\}$, which are the principle, orbital, and azimuthal numbers respectively that satisfy $n \geq \ell+1, \ell \geq 0, |m| \leq \ell$. The growth rates of the eigenstates scale very sensitively as $ \Gamma \propto \alpha^{4\ell + 5} / M$~\cite{Baumann:2019eav, Detweiler:1980uk, Dolan:2007mj}, where $\alpha$ is the gravitational fine structure constant
\begin{equation}
    \alpha \equiv \frac{M \mu}{M_{\rm pl}^2} \simeq 0.23 \left( \frac{M}{3 M_\odot} \right) \left( \frac{\mu}{10^{-11} \text{ eV}} \right) \, , 
\end{equation}
which is similar to $\alpha_{\rm BS}$ in (\ref{eqn:mini_bosonstar_Mmax}) for the boson star except that here $M$ is the black hole mass while $\alpha_{\rm BS}$ is defined with respect to boson star mass. The first few fastest-growing eigenstates are therefore the $\ket{n \ell m} = \ket{211}, \ket{322}$, and $\ket{433}$ modes (listed in decreasing order) and, depending on the value of $\alpha$, could grow well within astrophysical timescales. Superradiant amplification would occur when $\alpha / \ell \lesssim 0.4$, where the upper bound is saturated when the black hole is initially spinning maximally. Consequently, the higher the $\ell-$eigenstate the wider the range of $\alpha$ probed. 

\vskip 4pt

When the gravitational atom is part of a binary system, various resonance phenomena would be excited which could significantly backreact onto the orbital dynamics~\cite{Baumann:2018vus, Chia:2020dye, Baumann:2019ztm, Zhang:2018kib,  Ding:2020bnl, Baumann:2021fkf, Su:2021dwz, Takahashi:2023flk}. However, these resonances typically occur when the binary companion is either near or within the Bohr radius of the cloud, $r_c$ -- when the orbital separation is larger than $r_c$, the cloud's tidal deformation remains the leading-order finite-size effect~\cite{Baumann:2018vus}. In the $\alpha / \ell \ll 1$ regime, the Love numbers of these boson clouds are~\cite{Baumann:2018vus, DeLuca:2021ite}
\begin{equation}
    \Lambda_{c, \ell} \simeq 3.3  \times 10^5 \hskip 2pt k_{c, \ell} \hskip 1pt \left( \frac{M_c/M}{0.1}\right) \left( \frac{0.3}{\alpha} \right)^8 \, , \quad  \label{eqn:atom_Lambda}
\end{equation}
where the constant $k_{c, \ell}$ varies with $\ell$ and $M_c$ is the mass of the cloud.\footnote{From dimensional analysis, $\Lambda_c \sim (M_c / M) (r_{c}/M)^4$~\cite{Baumann:2018vus}, where $r_c / M \simeq n^2 /\alpha^2$ is the Bohr radius of the boson cloud. We therefore expect the relative ratios between $k_{c, \ell=1} : k_{c, \ell=2} : k_{c, \ell=3}$ to be approximately $1:25.6:256$. This simple scaling is in approximate agreement with the more precise calculation performed for these coefficients, $k_{c, \ell=1} = 1$ and $k_{c, \ell=2} = 16$~\cite{DeLuca:2021ite}. Since the precise coefficient for $k_{c, \ell=3}$ was not shown explicitly in that work, we will use $k_{c, \ell=3} \approx 200$ as an order of magnitude approximation. \label{footnote:atom_ell} } Depending on the initial black hole spin, the superradiance process can in principle extract up to $29 \%$ of the initial black hole mass~\cite{Penrose:1969pc}. 
From (\ref{eqn:atom_Lambda}) we see that the Love number of the gravitational atom also scales very sensitively with $\alpha$ and can therefore easily span over several orders of magnitude.

\vskip 4pt

On the right panel of Fig.~\ref{fig:ultralight_dm} we show the parameter space of the scalar gravitational atom to which our rate constraints apply. Since these boson clouds can only form around black holes, we assume they are the heavier component of the binary (blue region in Fig.~\ref{fig:param_bound}). There we show the constraints applied specifically to the $\ket{433}$ mode because the $\alpha / \ell \lesssim 0.4$ condition implies that the $\ket{433}$ mode admits a wider feasible range of $\alpha$. For other lower $\ell-$eigenstates, the values on $\mu$ constrained would only change slightly by $k^{1/8}_{c, \ell} \sim \mathcal{O}(1)$ (see Footnote~\ref{footnote:atom_ell}) and the range at which $\alpha$ would be valid is reduced. Note that the ultralight boson masses that are probed by the gravitational atom are of a similar scale to those of the mini boson star --- this is the case because their respective physically feasible regimes in parameter space, $\alpha_{\rm BS} < 0.633$ and $\alpha / \ell \lesssim 0.4$, are of the same order of magnitude.

\vskip 4pt

Strictly speaking, the constraints shown on the right of Fig.~\ref{fig:ultralight_dm} only applies to the complex scalar field. For the real scalar field, a key consideration in interpreting this figure is that the real boson cloud consists of an oscillating quadrupole structure and therefore sources continuous gravitational waves~\cite{Arvanitaki:2010sy, Yoshino:2014}. The typical lifetimes of the real scalar cloud eigenstates are
\begin{equation}
    M_c(t) = \frac{M_{c, 0}}{1 + t / \tau_c} \, , \quad \tau_c \sim \tau_{c, \ell} \left( \frac{M}{3 M_\odot} \right) \left( \frac{0.3}{\alpha} \right)^{4\ell + 11} \, , 
\end{equation}
where $M_{c, 0}$ is the initial cloud mass, $\tau_{c, \ell} \sim 10^{-3}, 10^3, 10^9$ years for the $\ell=1, 2, 3$ modes respectively~\cite{Yoshino:2014}.\footnote{
    Note that these values are strictly derived in the $\alpha \ll 1$ regime. For larger values of $\alpha$ nonlinear effects would reduce the emission power and can therefore increase $\tau_{c, \ell}$ by up to two orders of magnitude~\cite{Yoshino:2014}. 
} In the $\alpha \gtrsim 0.3$ regime, the emission power for the real $\ket{211}$ and $\ket{322}$ eigenstates would be sufficiently high for the cloud to fully dissipate within a Hubble time, making them unlikely to have been part of a merging binary as observed by the LIGO and Virgo observatories. On the other hand, the $\ket{433}$ eigenstate (and higher-order $\ell-$eigenstates) could survive over a Hubble time and therefore be probed by ground-based detectors. Such finite-lifetime considerations do not apply to complex scalar fields because they possess conserved $U(1)$ charges and therefore do not emit continuous gravitational waves.

\subsubsection{Particle Dark Matter} \label{sec:particle_dm}

In this section we focus on \textit{massive boson stars} as a candidate compact object that is formed from particle dark matter. This is a rather rare example as compared to other BSM compact objects in that many of its macroscopic properties, including their Love numbers, have been investigated comprehensively in the literature. 

\subsubsection*{Massive Boson Stars} 

Massive boson stars~\cite{Schunck:2003kk, Colpi1986, Liebling:2012fv} are generalizations of the mini boson star investigated in \S\ref{sec:ultralight_dm} whereby additional nonlinear scalar self-interaction operators are included in the action. Although several classes of nonlinear couplings have been investigated in the literature, in this work we focus on the quartic interaction~\cite{Colpi1986} which is a natural renormalizable operator from an effective field theory point of view which preserves the global $U(1)$ symmetry. The action is
\begin{equation}
S = \int d^4 x \sqrt{-g} \left[ \frac{R}{16\pi} - \nabla_\alpha \Phi^* \nabla^\alpha \Phi - \mu^2 |\Phi|^2 - \frac{\lambda}{2} |\Phi|^4 \right] \, ,   \label{eqn:massive_BS_action}
\end{equation}
where we focus on the repulsive $\lambda > 0$ potential to ensure the Hamiltonian is bounded from below, $R$ is the Ricci scalar and $\mu$ is the boson mass. As we shall see, the presence of scalar nonlinearities introduces a new scale in solutions of boson stars. For stellar mass boson stars this new scale allows us to probe values of $\mu$ that are much larger than the $\sim 10^{-11}$eV range for mini boson stars.

\vskip 4pt

In general, solutions to (\ref{eqn:massive_BS_action}) for arbitrary values of $\lambda$ have to be obtained numerically. Remarkably, various analytic results can be derived for the massive boson star when the quartic coupling is large compared to the boson mass in Planck units, $\lambda \gg ( \mu / M_{\rm pl})^2$. In fact, this so-called ``strong-coupling limit"~\cite{Colpi1986} is naturally satisfied by most dark matter scenarios: dark matter particles are often postulated to have masses that are smaller than the Planck mass, while technical naturalness arguments typically demand $\lambda \sim \mathcal{O}(1)$. As we shall see below, solar-mass boson stars would only form if the particle mass lies within $\mu \lesssim \mathcal{O}(100\text{MeV})$; the large hierarchy $(\mu / M_{\rm pl})^2 \lesssim 10^{-40}$ therefore makes the strong-coupling limit trivially satisfied in most models. As is conventional in the boson star literature, we define the rescaled quartic coupling $\tilde{\lambda} \equiv \lambda \hskip 1pt ( M_{\rm pl} / \mu)^2 / 8 \pi$ such that the strong coupling limit is equivalent to $\tilde{\lambda} \gg 1$.

\begin{figure}[t!]
    \centering
    \includegraphics[width=0.6\textwidth, trim=0 0 0 0]{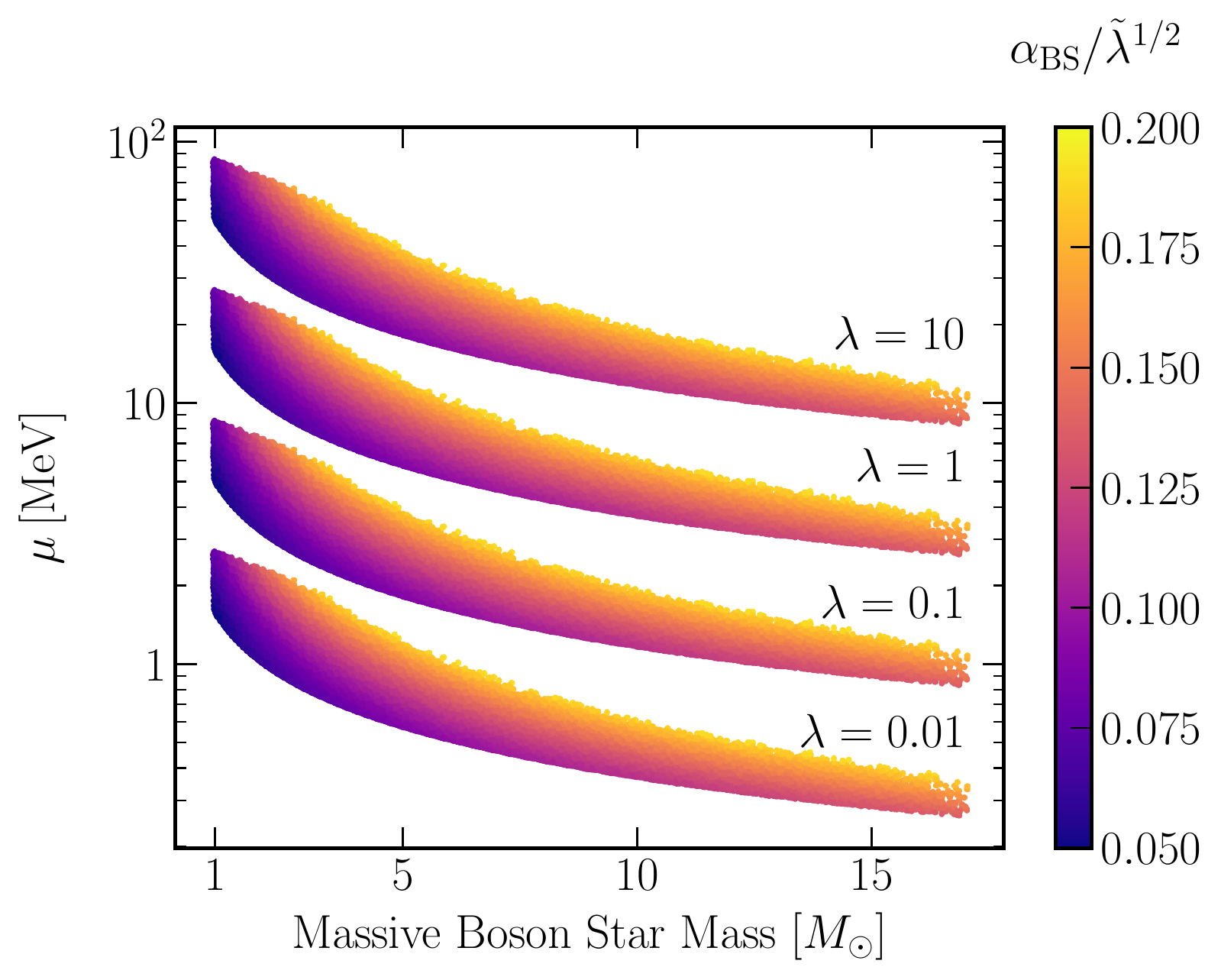}
    \caption{Same as the left panel of Fig.~\ref{fig:ultralight_dm}, but instead translating our constraints to boson stars (BS) with relatively massive boson particles and including quartic self-interactions; see (\ref{eqn:massive_BS_action}). Each colored band corresponds to a different value of the quartic-interaction parameter $\lambda$; for fixed $\alpha_{\rm BS} / \tilde{\lambda}^{1/2}$ the relative vertical scale between the bands scale with the boson mass as $\mu \propto \sqrt{\lambda}$. The physically feasible regime for the massive boson star lies in $\alpha_{\rm BS} /  \tilde{\lambda}^{1/2} \lesssim 0.22$. } 
    \label{fig:particle_dm}
\end{figure}

\vskip 4pt

In the limit where $\tilde{\lambda} \to \infty$ but $\lambda$ remains finite, the maximum mass of the boson star is~\cite{Colpi1986}
\begin{equation}
    M_{\rm BS, max} \approx 0.044 \sqrt{\lambda} \hskip 2pt \frac{M_{\rm pl}^3}{\mu^2} = 7.2 M_\odot \sqrt{\lambda} \left( \frac{100 \text{ MeV}}{\mu} \right)^2 \, . \label{eqn:massive_bosonstar_Mmax}
\end{equation}
Unlike the mini boson stars, whose maximum mass scale as $M_{\rm BS, max} \sim M_{\rm pl}^2 / \mu$, cf. (\ref{eqn:mini_bosonstar_Mmax}), here the maximum mass of the massive boson star scales in a similar way to the Chandrasekhar limit for neutrons stars. 
This is the case because the repulsive potential provides an additional source of pressure against the star's self-gravity, thereby allowing the bosonic particle to be more massive than that is the mini boson star. 
The Love numbers of the massive boson stars have also been computed in Ref.~\cite{Sennett:2017etc} for a range of values of $\tilde{\lambda}$. In the $\tilde{\lambda} \to \infty$ and finite-$\lambda$ limit, the Love number of the boson star is approximately
\begin{equation}
    \Lambda_{\rm BS} \simeq 300 \left( \frac{0.22}{\alpha_{{\rm BS}} \hskip 2pt / \hskip 2pt  \tilde{\lambda}^{1/2}} \right)^{10} \, , \quad \frac{\alpha_{\rm BS}}{\tilde{\lambda}^{1/2}} \equiv \frac{ M_{\rm BS} \mu}{M^2_{\rm pl} \hskip 2pt \tilde{\lambda}^{1/2}} \approx 0.22 \left( \frac{M_{\rm BS}}{M_{\rm BS, max}}\right) \, , \label{eqn:massive_BS_Lambda}
\end{equation}
where we extract the overall coefficient and the parametric scaling in $\alpha_{{\rm BS}} / \tilde{\lambda}^{1/2}$ from the right panel of Fig.~2 in~\cite{Sennett:2017etc}. Similar to the mini boson star case in (\ref{eqn:mini_BS_Lambda}), the $ \Lambda_{\rm BS} \propto \alpha_{{\rm BS}}^{-10}$ scaling arises from the fact that the stellar radius scales as $r_{\rm BS}/m \sim \alpha_{{\rm BS}}^{-2}$. Crucially, unlike the mini boson star, the Love numbers of massive boson stars are proportional to $\sqrt{\lambda}$ and the physical range of parameters lie within $\alpha_{\rm BS} / \tilde{\lambda}^{1/2} \lesssim 0.22$.

\vskip 4pt

In Fig.~\ref{fig:particle_dm} we use (\ref{eqn:massive_BS_Lambda}) to translate the constraints of our search space into the parameter space of the massive boson star. We choose a few representative values of $\lambda$, illustrating how each ``band" for a given value of $\lambda$ would shift on the vertical scale as $\mu \propto \sqrt{\lambda}$ . Crucially, the boson mass scale that is constrained for this compact object is $\lesssim \mathcal{O}(100\text{MeV})$, which is orders of magnitude larger than that of the mini boson star in Fig.~\ref{fig:ultralight_dm}. In principle one could constrain boson masses below the MeV scale by considering $\lambda \ll 10^{-2}$ while maintaining the strong coupling condition, though such small $\lambda$ couplings generally demand the existence of an underlying (approximate) symmetry that is a priori absent in (\ref{eqn:massive_BS_action}); see e.g. Ref.~\cite{AmaroSeoane:2010qx} for constraints applied to that region of parameter space from astrophysical considerations.

\pagebreak

\section{Conclusions and Outlook} \label{sec:conclusions}

In this paper, we conducted the first matched-filtering search for compact objects with large Love numbers in the O1--O3 LIGO Hanford and Livingston data. We first construct a PN inspiral-only waveform model for objects with large tidal deformabilities and build an effectual template bank (Figs.~\ref{fig:phase_derivative}, \ref{fig:Aref_effectualness}). Our search spans effective binary tidal deformabilities $10^2 \lesssim \tilde{\Lambda} \lesssim 10^6$ and chirp masses $3 M_\odot < \mathcal{M} < 15 M_\odot$, cf. Fig.~\ref{fig:bank_params} (with the lower bound of $\tilde{\Lambda}$ intentionally chosen to avoid the $\tilde{\Lambda} = 0$ BBH value). We list the detection statistics of our top three triggers with large Love numbers in Table~\ref{tab:signalsFound_new}, concluding that they are not statistically significant. In Fig.~\ref{fig:ifar} we illustrate the IFAR distribution for all observing runs, demonstrating how our triggers fall within the $\sim 1-2\sigma$ range of Poisson noise.

\vskip 4pt

We used our null detection to place an upper limit on the merger rates of such binary systems in the LIGO and Virgo bands (Fig.~\ref{fig:vt}). This upper limit is model-agnostic and applies to any type of compact object whose masses and Love numbers fall within the parameter space indicated in Fig.~\ref{fig:param_bound}, and whose spins are negligible. 
We discuss the implications of these constraints on BSM physics, including several scenarios for ultralight boson field and particle dark matter (Figs.~\ref{fig:ultralight_dm}, \ref{fig:particle_dm}). While we focused on boson stars~\cite{Kaup1968, Ruffini1969, Breit:1983nr, Jetzer:1991jr, Colpi1986, Schunck:2003kk,Liebling:2012fv} and gravitational atoms~\cite{Arvanitaki:2009fg, Arvanitaki:2010sy, Baumann:2019eav, Detweiler:1980uk, Dolan:2007mj}, our model-independent rate constraint applies to any BSM compact object as long as their orbital dynamics during inspiral are captured by our tidal waveform. Note that the merger rate constraints can also be used to constrain the abundance of exotic objects and the fraction of dark matter that could be composed of these objects -- we hope to pursue this question in future work.

\vskip 4pt

Remarkably, using the non-spinning inspiral-only tidal waveform we are able to recover many of the known BBH events that were previously identified using full IMR searches~\cite{LIGOScientific:2018mvr, LIGOScientific:2020ibl, LIGOScientific:2021usb, LIGOScientific:2021djp, Venumadhav:2019tad, Venumadhav:2019lyq, Olsen:2022pin, Ajit-O3b, Nitz:2018imz, Nitz:2019hdf, Nitz:2021uxj, Nitz:2021zwj}. The detection statistics for our triggers from these events are listed in Table~\ref{tab:signalsFound}. It is worth noting that all of the best-fit templates in that table have $\tilde{\Lambda} \sim 10^2 - 10^3$. This is not surprising because, as we have demonstrated in Fig.~\ref{fig:effectualness}, these relatively small values of $\tilde{\Lambda}$ are consistent with the $\tilde{\Lambda}=0$ BBH value~\cite{Chia:2020yla, Charalambous:2021mea, Binnington:2009bb, Damour:2009vw, Kol:2011vg} at the level of a matched-filtering search. Interestingly, almost half of the previously-identified BBHs, which are labeled by the ${}^\dagger$ superscript in Table~\ref{tab:signalsFound}, have masses that fall outside of our search parameter space. While the SNRs of all of the inspiral-only BBH triggers are lower than those reported in full IMR searches~\cite{LIGOScientific:2018mvr, LIGOScientific:2020ibl, LIGOScientific:2021usb, LIGOScientific:2021djp, Venumadhav:2019tad, Venumadhav:2019lyq, Olsen:2022pin, Ajit-O3b, Nitz:2018imz, Nitz:2019hdf, Nitz:2021uxj, Nitz:2021zwj}, our work is the first to demonstrate the ability of using an inspiral-only waveform to search for black holes. Furthermore, our work has the added advantage of easily allowing for additional effects to be incorporated into the waveform model as needed --- a desirable quality when searching for new signals.

\vskip 4pt

In Fig.~\ref{fig:Love_constraint}, we show parameter estimation results for the Love numbers of black holes using the tidal waveform. We constrain the BBH effective tidal parameter to $\tilde{\Lambda} \lesssim 10^3$ at the $90\%$ credible interval, which is broadly consistent with the results reported in Ref.~\cite{Narikawa:2021pak}. Future detectors will potentially improve this constraint by at least an order of magnitude~\cite{Cardoso:2017cfl, Puecher:2023twf}. In Fig.~\ref{fig:corner}, we additionally show that there is a mild degeneracy between $\chi_{\rm eff}$ and $\tilde{\Lambda}$. This degeneracy, however, did not help in recovering most of the known BBH merger signals, which have positive spins~\cite{LIGOScientific:2021usb, LIGOScientific:2021djp}, because the degeneracy direction implies that $\tilde{\Lambda} > 0$ would partially mimic $\chi_{\rm eff} < 0$. Finally, in Fig.~\ref{fig:violin} we show that using an inspiral-only waveform for parameter estimation leads to small biases in the recovered parameters compared to more advanced IMR waveforms such as \texttt{IMRPhenomXPHM}. This bias comes from a combination of our prior support vanishing at $\tilde{\Lambda}=0$ together with an incomplete set of phase terms at high PN order. Pushing analytic waveform development to higher PN orders will therefore be highly beneficial for future analyses.

\vskip 4pt

This work represents the first-ever dedicated matched-filtering search that extends beyond previous efforts where templates were limited to detecting black holes and neutron stars. However, we believe we have barely scratched the surface of the boundless potential for new discoveries in this emerging era of GW astronomy. Our analysis can be extended in a variety of ways:

\begin{itemize}

  \item \textit{Low-mass binary systems:} We restrict ourselves to the $\mathcal{M} > 3 M_\odot$ BBH mass range because the number of templates required for an effectual template bank, and therefore the computational cost of the search, increases rapidly as one extends the lower bound of the mass range into the neutron star regime and below. 
  However, low-mass binaries are the most well-motivated part of the parameter space because \textit{i)} the range of $\tilde{\Lambda}$ which could be probed by our template bank significantly widens as the binary mass decreases (see Fig.~\ref{fig:bank_params}); and \textit{ii)} the masses of compact objects in many BSM scenarios are often bounded from above, see e.g. (\ref{eqn:mini_bosonstar_Mmax}) and (\ref{eqn:massive_bosonstar_Mmax}) for boson stars, but otherwise admit stable solutions at lower masses. A dedicated search in the low-mass and large-$\tilde{\Lambda}$ region is therefore not only interesting observationally, but also highly-motivated theoretically;

    \item \textit{Additional finite-size effects:} We modeled the compact objects as point particles with large tidal deformabilities, which is an excellent starting point since the tidal deformability is the leading non-spinning finite-size effect. However, other finite-size effects, such as spin-induced moments~\cite{Hansen:1974zz, Thorne:1980ru, Poisson:1997ha, Krishnendu:2017shb} and tidal dissipation~\cite{poisson_will_2014, Chia:2020yla, Charalambous:2021mea, Isoyama:2017tbp, Goldberger:2020fot}, would also generically be present in binaries with exotic compact objects. A comprehensive search should consider all of the above effects in order to more accurately model the full space of potential signals. However, doing so would come at a price of including more free parameters, increasing the dimensionality of the template bank and ultimately increasing the search's computational cost.
    Refs.~\cite{Coogan:2022qxs, Chia:2022rwc} have taken the first steps towards a search similar to the one reported here, for objects with large spin-induced quadrupole moments;

    \item \textit{Other physical imprints:} In addition to other finite-size effects, our templates do not incorporate various BSM imprints which cannot be mimicked by the 5PN tidal parameter $\tilde{\Lambda}$. These include, but are not limited to, the resonances excited in gravitational atoms when their binary companion approaches the Bohr radii~\cite{Baumann:2018vus, Chia:2020dye, Baumann:2019ztm, Zhang:2018kib,  Ding:2020bnl, Baumann:2021fkf, Su:2021dwz, Takahashi:2023flk}, an additional long-range fifth force induced in binary neutron stars~\cite{Hook:2017psm, Huang:2018pbu, Brax:2021qqo}, dynamical effects generated in modified theories of gravity~\cite{Yagi:2011xp, Silva:2020omi, Okounkova:2020rqw, Shiralilou:2020gah, Shiralilou:2021mfl, East:2022rqi}, and so on. Dedicated searches for these effects, if they are sufficiently large to deviate from BBH waveforms, would require precise additional modeling;

    \item \textit{Full IMR model-dependent waveforms:} We focused on the inspiral portion of the signal because the imprints of new physics can be incorporated analytically in this regime, thereby allowing for a source-agnostic search for exotic compact objects that does not require specifying a model for the strong-field physics determining their merger signals. It would be interesting to perform searches using full IMR waveforms for specific BSM compact objects, see e.g.~Refs.~\cite{Bezares:2018qwa, Bezares:2022obu, Croft:2022bxq, Bamber:2022pbs, Siemonsen:2023hko}, as this would increase the statistical significance of putative detections. Such a search, however, would require waveform calibration between the analytic inspiral and full numerical relativity simulations, and 
    its scope would be limited to to the specific type of exotic objects being modeled (i.e. losing the source-agnostic advantage);

    \item \textit{Space-based GW observations:} Future space-based GW detectors which observe the $\sim 10^{-4} - 1$ Hz frequency range, including LISA~\cite{2017arXiv170200786A}, TianQin~\cite{Luo:2015ght}, Taiji~\cite{Guo:2018npi}, MAGIS~\cite{Graham:2017pmn} and DECIGO~\cite{Kawamura:2011zz}, would probe new binaries over a vast
    region of unexplored parameter space. Following the same argument which led to (\ref{eqn:lambda_bound}), space-based detectors would in principle probe objects with compactness and Love numbers over the ranges
    \begin{equation}
    \frac{r}{m} \lesssim 2 \times 10^5 \left( \frac{10^{-3} \text{ Hz}}{f_{\rm low}} \right)^{2/3} \left(\frac{ M_\odot}{m} \right)^{2/3} \, , \quad \Lambda  \lesssim  10^{26} \left( \frac{10^{-3} \text{ Hz}}{f_{\rm low}} \right)^{10/3}  \left(\frac{ M_\odot}{m} \right)^{10/3}  \, , \label{eqn:lambda_bound_lisa}
    \end{equation}
    which is significantly wider than the reference values shown in (\ref{eqn:lambda_bound}) for ground-based detectors. Future template-based searches must therefore accurately incorporate the imprints of large Love numbers (and other physical effects) in waveform models to make the most of the discovery potential of space-based GW astronomy.

\end{itemize}

\noindent We hope to pursue these research directions in future work. The pursuit of Love, and other forces of Nature, shall carry on.

\begin{acknowledgments}
We thank Mustafa Amin, Katerina Chatziioannou, Adam Coogan, Richard George, Tanja Hinderer, Max Isi,  Misha Ivanov, Marc Kamionkowski, Cody Messick, Ani Prabhu, Carolyn Raithel, Nashwan Sabti, and Kaze Wong for stimulating discussions. 
HSC gratefully acknowledges support from the Institute for Advanced Study.
DW gratefully acknowledges support from the Friends of the Institute for Advanced Study Membership and from the W. M. Keck Foundation Fund.
AZ is supported by NSF Grants PHY-2207594 and PHY-2308833. T.E. is supported by the Horizon Postdoctoral Fellowship. TV acknowledges support from NSF Grants PHY-2012086 and PHY-2309360, the Alfred P. Sloan Foundation, and the Hellman Family Faculty Fellowship.
This paper has been assigned preprint numbers LIGO-P2300154 and UTWI-16-2023.
Finally, we acknowledge the use of the Python modules \texttt{jupyter}~\cite{jupyter}, \texttt{matplotlib}~\cite{Hunter:2007}, \texttt{numpy}~\cite{numpy}, \texttt{scipy}~\cite{scipy}, \texttt{scikit-learn}~\cite{DBLP:journals/corr/BuitinckLBPMGNPGGLVJHV13} and \texttt{tqdm}~\cite{tqdm}.

\vskip 8pt

This research has made use of data, software and/or web tools obtained from the Gravitational Wave Open Science Center (\href{https://www.gw-openscience.org/}{https://www.gw-openscience.org/}), a service of LIGO Laboratory, the LIGO Scientific Collaboration and the Virgo Collaboration. LIGO Laboratory and Advanced LIGO are funded by the United States National Science Foundation (NSF) as well as the Science and Technology Facilities Council (STFC) of the United Kingdom, the Max-Planck-Society (MPS), and the State of Niedersachsen/Germany for support of the construction of Advanced LIGO and construction and operation of the GEO600 detector. Additional support for Advanced LIGO was provided by the Australian Research Council. Virgo is funded, through the European Gravitational Observatory (EGO), by the French Centre National de Recherche Scientifique (CNRS), the Italian Istituto Nazionale di Fisica Nucleare (INFN) and the Dutch Nikhef, with contributions by institutions from Belgium, Germany, Greece, Hungary, Ireland, Japan, Monaco, Poland, Portugal, Spain.
\end{acknowledgments}

\newpage
\phantomsection
\addcontentsline{toc}{section}{References}
\bibliographystyle{utphys}
\bibliography{references.bib}
\end{document}